\documentclass[prb,preprint,aps]{revtex4}
\usepackage{epsfig}
\usepackage{amsmath}
\usepackage{amssymb}
\usepackage{graphicx}
\usepackage{multirow}
\usepackage[usenames]{color}
\usepackage[english]{babel}

\def\Im{\hbox{Im}}

\newlength{\bxwidth}

\newlength{\spear}

\newlength{\figwidth}
\newcommand{\fg}[3]
{\begin{figure}[tb]
{\includegraphics[width=\figwidth]{#1}}
\caption{#2}\label{#3}\end{figure}}

\def\gtappr{{{\lower4pt\hbox{$>$} } \atop \widetilde{ \ \ \ }}}
\def\ltappr{{{\lower4pt\hbox{$<$} } \atop \widetilde{ \ \ \ }}}

\newlength{\upit}\upit=0.1truein

\newlength{\wx}
\newlength{\wy}
\newlength{\wz}
\begin{document}
\title{Quantum Critical Paraelectrics and the Casimir Effect in Time}         
\author{L. P\'alov\'a, P. Chandra, P. Coleman}
\affiliation{Center for Materials Theory, Department of 
Physics and Astronomy, Rutgers University, 
Piscataway, NJ 08854}
\date{\today}
\begin{abstract}
We study the quantum paraelectric-ferroelectric transition near a 
quantum critical point, emphasizing the role of temperature as a
``finite size effect'' in time.  The influence of 
temperature near quantum criticality may thus be likened to 
a temporal Casimir effect.
The resulting finite-size scaling approach yields
$\frac{1}{T^2}$ behavior of the paraelectric susceptibility
($\chi$) and the scaling form 
$\chi(\omega,T) = \frac{1}{\omega^2} F(\frac{\omega}{T})$,
recovering results previously found by more technical
methods. We use a Gaussian theory to illustrate how these 
temperature-dependences emerge from a microscopic approach; we characterize
the classical-quantum crossover in $\chi$, and the resulting
phase diagram is presented.  We also show that coupling to an acoustic 
phonon at low temperatures ($T$) is relevant and influences the 
transition line, possibly resulting in a reentrant quantum ferroelectric
phase. Observable consequences of our approach for 
measurements on specific paraelectric materials at low temperatures
are discussed.
\end{abstract}
\maketitle
\section{Introduction}
The role of temperature in the vicinity of a quantum phase transition
is distinct from that close to its classical counterpart, where it acts as 
a tuning parameter. Near a quantum critical point (QCP), temperature 
provides a low energy cut-off for quantum fluctuations; the 
associated {\sl finite} time-scale is defined through the uncertainty 
relation $\Delta t  \sim \frac{\hbar }{k_{B}T}$.
This same phenomenon manifests itself as a boundary condition
in the Feynman path integral; it is in this sense that temperature plays the
role of a {\sl finite-size effect in time} at a quantum critical point.
\cite{Cardy96,Sondhi97,Sachdev99,Continentino01,Coleman05}
The interplay between the scale-invariant quantum critical fluctuations and 
the temporal boundary condition imposed by temperature 
is reminiscent of the Casimir effect,\cite{Casimir48,Krech94,Kardar99}  
where neutral metallic structures attract each other 
\cite{Lamoreaux97,Mohideen98,Chan01,Lisanti05,Obrecht07} due to 
zero-point vacuum fluctuations.  

In this paper we explore the observable ramifications
of temperature as a temporal Casimir effect, applying it
to the example of a quantum ferroelectric critical point
(QFCP) where detailed interplay between theory and experiment is
possible below, at and above the upper critical 
dimension. Our work is motivated by recent experiments on the 
quantum paraelectric $SrTiO_3$ (STO) where $1/T^{2}$ behavior is 
measured in the dielectric susceptibility 
near the QFCP.\cite{Venturini04,Coleman06,Rowley07}
Here we show how this result is simply obtained  using finite-size 
scaling in time; more generally we present similar derivations of 
several measurable quantities, 
recovering results that have been previously derived using 
more technical diagrammatic,\cite{Rechester71,Khmel'nitskii71,Roussev03,Das07} 
large $N$ \cite{Schneider76} and renormalization group 
methods.\cite{Schmeltzer83,Sachdev97}
In particular we present a simple interpretation of finite-temperature 
crossover functions near quantum critical points 
previously found using $\epsilon$-expansion 
techniques,\cite{Sachdev97} and link them to ongoing low-temperature 
experiments on quantum paraelectric materials.  
We illustrate these ideas using a Gaussian theory to characterize
the domain of influence of the QFCP and we present the full phase diagram. 
Next we expand upon previous work by tuning away 
from the QFCP, studying deviations from scaling;
here we find that coupling between the soft polarization and
long-wavelength acoustic phonon modes is relevant and can lead to
a shift of phase boundaries and to a reentrant quantum 
ferroelectric phase.  We end with a discussion of our results
and with questions to be pursued in future work.

\section{The Casimir Effect in Space and in Time}

The Casimir effect is a boundary condition response
of the electromagnetic vacuum.  
The gapless nature of the photon spectrum
means that the zero-point electromagnetic fluctuations are scale-invariant; the
vacuum is literally in a quantum critical state. However, once the
boundary conditions are introduced, the system is tuned away from
criticality and develops a finite correlation length, $\xi$.
The Coulomb interaction between two charges, the correlation 
function of the electromagnetic
potential inside the cavity, is changed from the vacuum to the cavity as   
\begin{equation}
V(q)_{free} \sim \langle \delta \phi_q \delta \phi_{-q} \rangle = \frac{e^2}{ q^2} 
\qquad \rightarrow \qquad
V(q)_{cavity} \sim  \frac{e^2} {q_{\perp}^2+ \xi^{-2} } \qquad\qquad 
\xi = \frac{a}{\pi}
\label{Vcavity}
\end{equation}
where the plates have removed field modes and have introduced a finite 
$\xi$. 
In an analogous way, the 
partition function of a quantum system at finite temperatures
is described by a Feynman path integral over the 
configurations of the fields in Euclidean space-time~\cite{Hertz73}
where temperature introduces a cutoff in the temporal direction.   In Figure
1 we present a visual comparison of the Casimir effect in space and in
time. In both cases, the finite boundary effects induce the replacement of
a continuum of quantum mechanical modes by a discrete spectrum of excitations.

\figwidth=5in
\fg{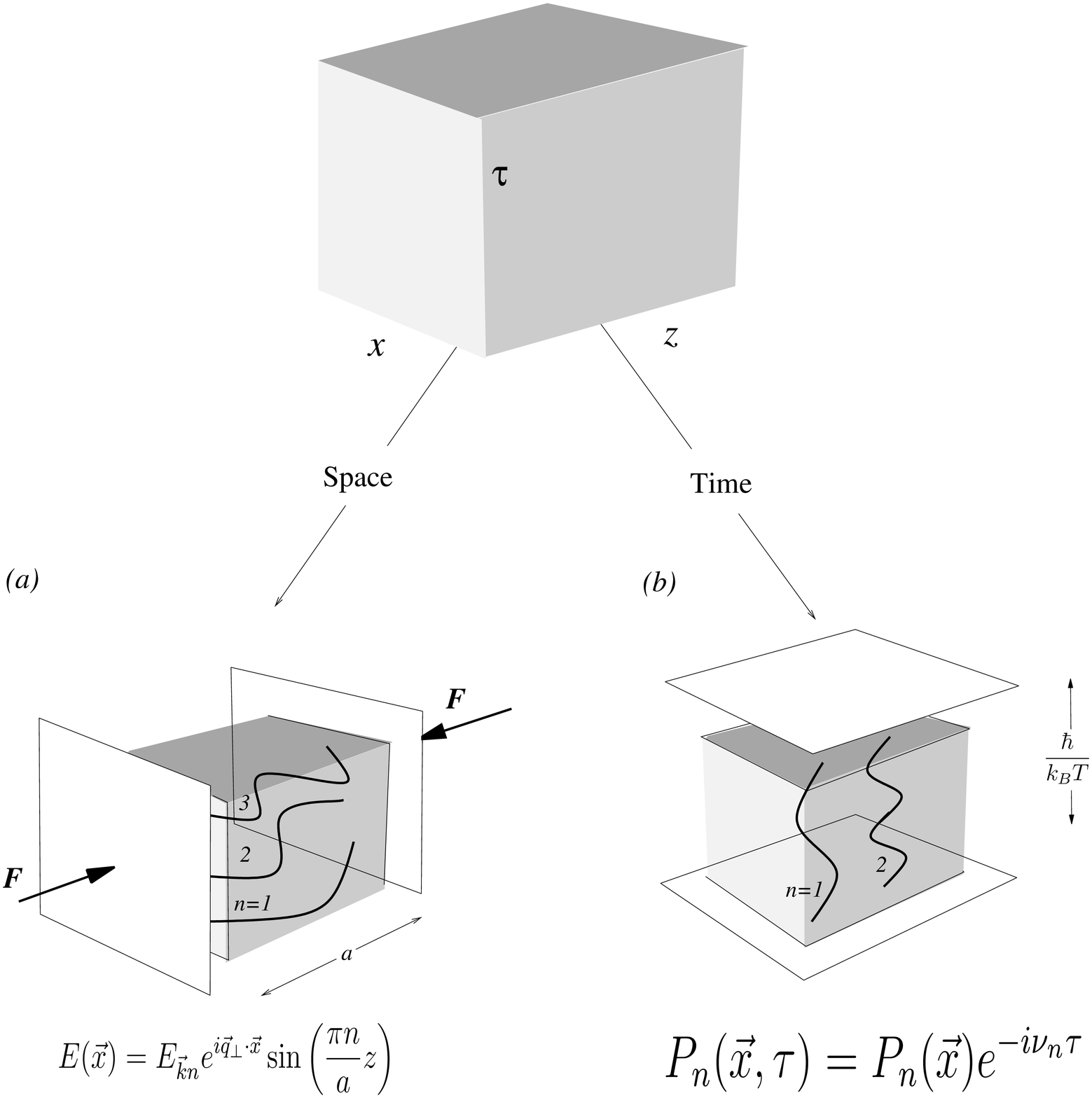}
{Casimir effect in space and time. (a) Imposition of spatial
boundaries on the quantum critical electromagnetic field yields the
conventional Casimir effect (b) Imposition of temporal boundary
conditions on a quantum critical paraelectric generates the effect of
non-zero temperature. }
{fig1}

In the quantum paraelectric of  interest here, the 
path integral is taken over the space-time configurations of the polarization 
field $P (\vec{x},\tau )$, 
\begin{equation}
Z = \sum_{{\{ P (x,\tau ) \}}} \exp \left[- \frac{S_E [P]}{\hbar} \right],
\end{equation}
where
\begin{equation}
S_E [P] = \int_{0}^{\frac{\hbar }{k_{B}T}} d\tau d^{3}x \, {\cal L}_E[P]
\label{SEuclidean}
\end{equation}
and ${\cal L}_E[P]$ is the Lagrangian in Euclidean space-time. 
The action per unit time is now the
Free energy $F$ of the system (See Table I.).
The salient point is that finite temperature
imposes a {\sl boundary condition in imaginary time} and the allowed
configurations 
of the bosonic quantum fields are periodic in the imaginary time
interval $\tau \in [0, \hbar \beta] $ ($\beta \equiv \frac{1}{k_{B}T}$) so that
$\vec{P} (\vec{x},\tau )= \vec{P} (\vec{x}, \tau + \hbar \beta )$,
 which permits the 
quantum fields are thus decomposed in terms  of a discrete
set of Fourier modes 
\begin{equation}
P_{n} (\vec{x},\tau )=\sum_{\vec{q},n} 
P (\vec{q}, i\nu_n)  
e^{i (\vec{q}\cdot \vec{x}-  \nu_n \tau )}
\end{equation}
where
\begin{equation}
\nu_n= n \left(\frac{2 \pi k_B T }{\hbar}\right)
\label{nu}
\end{equation}
are the discrete Matsubara frequencies; we recall that at $T=0$ the (imaginary) frequency
spectrum is a continuum.
The response and correlation functions in (discrete) imaginary frequency
\begin{equation}
\chi_E(\vec{q},i\nu_n) =   \langle P(\vec{q},i\nu_n) P(-\vec{q},-i\nu_n)\rangle
\label{iresponse}
\end{equation}
can be analytically
continued to yield the retarded response function 
\begin{equation}
\chi_E(\vec{q}, i\nu_n) \rightarrow \chi_E(\vec{q}, \omega)= \chi_{E}
(\vec{q},z)\vert_{z = \omega+i\delta }
\end{equation}
where $\omega$ is a real frequency; for writing convenience we will subsequently
drop the ``E'' subscript in $\chi$ e.g. 
$\chi(\vec{q}, \omega) \equiv \chi_E (\vec{q}, \omega)$.

\centerline{\bf Table. 1. Casimir Effect and Quantum Criticality. }

\begin{center}
\label{Casimirtab}
\begin{tabular}{|l||c|c|}
\hline
\multirow{2}{15mm}{}&Casimir&Finite Temperature Effects Near\\
& Effect & Quantum Criticality\\
\hline
Boundary condition& Space 
& Time 
\\
\hline
``S matrix''& $U = e^{-i E{\rm  \bar{t}}/\hbar }$
& $Z= e^{- \beta F}$
\\
\hline
\multirow{3}{15mm}&&\\
Path Integral& 
$\displaystyle U =\int D[\phi ] \exp\left[{ i \frac{S[\phi]}{\hbar }} \right]
$
&
$\displaystyle Z = \int D[P] \exp\left[{- \frac{S_{E}[P]}{\hbar }} \right]
$
\\
&&\\
\hline
\multirow{3}{10mm}&&\\
Action/time& 
$\displaystyle E
$
&
$\displaystyle \frac{S_{E}}{\beta \hbar } = F
$
\\
&&\\
\hline
Time interval& 
$\rm  \bar{t} (\rightarrow \infty )$
&
$\beta \hbar $
\\
\hline
Spatial interval & 
a 
&
$\infty $
\\
\hline
\multirow{3}{15mm}&&\\
Discrete wavevector/frequency
& 
$q_{zn}=\left(\frac{\pi}{a} \right)
n$
&
$\nu_{n} = \left(\frac{2\pi k_{B}T}{\hbar } \right)
n$
\\
&&\\
\hline
\end{tabular}
\end{center}

\label{Casimirtab}

Like the parallel plates in the traditional Casimir
effect, temperature removes modes of the field.  
In this case it is the frequencies not the
wavevectors that assume a discrete character, namely
\begin{equation}
q= (\vec{q},\omega)\rightarrow (\vec{q}, i \nu_n),
\end{equation}
where $\nu_n$ are defined in (\ref{nu}).
%
%

The Casimir analogy must be used with care. 
In contrast to the noninteracting nature of the low-energy
electromagnetic field, the modes at a typical QCP
are interacting.  In the conventional Casimir effect, the finite
correlation length is induced purely through the discretization of 
momenta perpendicular to the plates.
By contrast, at an interacting QCP, the discretization of
Matsubara frequencies imposed by the boundary condition generates the 
thermal fluctuations in the fields in real time.
These are fed back via interactions to generate a temperature-dependent 
gap in the spectrum
and a finite correlation time. 
Despite the complicated nature of this feedback, 
provided the underlying system is critical, temperature acting 
as a boundary condition in time 
will set the scale of the finite 
correlation time 
\begin{equation}
\xi_{\tau} = \frac{\hbar}{\kappa k_B T},
\label{ltau}
\end{equation}
where $\kappa$ is a constant. In cases where the
quantum critical physics is universal, such as ferroelectrics
in dimensions below $d=3$, 
we expect the coefficient $\kappa$ to be also universal and independent of the
underlying strength  of the mode-mode coupling. 
The ``temporal confinement'' of the fields in imaginary time
thus manifests itself as a finite response time in the real-time 
correlation and response functions.

For the quantum
paraelectric at the QFCP, the imaginary time
correlation functions are scale-invariant
\begin{equation}
\chi (\vec{q},i\nu )=\left. \phantom{\int}\langle P(\vec{q},i\nu) P(-\vec{q},-i \nu) \rangle \right \vert_{T=0} 
\sim \frac{1}{\nu^2 + c_s^2 q^2}.
\label{PPQFCPA}
\end{equation}
At a finite temperature this response function
acquires a finite correlation time
\begin{equation}
\chi (\vec{q},i\nu_n)
\sim 
\frac{1}{\nu_n^2 + c_s^2q^2+ \xi_{\tau }^{-2}}
\label{PPQFCP}
\end{equation}
where
\begin{equation}
\xi_\tau^{-2} = 3\gamma_c \left\{ \langle P^2 \rangle_{T \neq 0} - \langle P^2 \rangle_{T=0} \right\}
\label{ltau2}
\end{equation}
is determined by mode-mode interactions, 
where $\gamma_c$ is the coupling constant describing the quartic
interactions between the modes,  to be defined in Sec IV.
We note, as shall be shown explicitly in Section IV, that for dimensions $d$
such that $1 < d < 3$, the feedback will be sufficiently strong
such that 
$\xi_\tau$ will be {\sl independent} of the coupling constant $\gamma_c$;
by contrast for $d > 3$ 
the
feedback effects are weak so that there will be a $\gamma_c$-dependence of
$\xi_\tau$. 
The case $d = d_c^u= 3$
is marginal and will be discussed as a distinct case. At a temperature
above a quantum
critical point, the energy scale 
\begin{eqnarray}\label{definegap}
\Delta (T)=  
{\alpha } k_{B}T
\end{eqnarray}
will set  the size of the gap in the phonon dispersion relation.
Here $\Delta(T) \sim \hbar \xi_\tau^{-1}$ and
$\alpha = O (1)$ is a constant of proportionality.

Reconnecting to our previous discussion, we remark that
real-time response
functions from expressions like (\ref{PPQFCP}) are obtained by analytic 
continuation to real frequencies 
$i\nu_{n}\rightarrow \omega$.
Since 
$\xi_{\tau} \sim \frac{1}{T}$,
the dielectric susceptibility
in the approach to the QFCP has the temperature-dependence
\begin{equation}
\chi (T)= \left . \phantom{\int}\chi (q,i\nu_n)\right\vert_{q=0, \nu=0} 
\sim \xi_{\tau }^{2}
\propto \frac{1}{T^2}
\label{chiCasimir}
\end{equation}
in contrast to the Curie form ($\chi \sim \frac{1}{T}$) associated with a classical paraelectric;
this $1/T^2$ temperature-dependence was previously derived from a diagrammatic
resummation,\cite{Rechester71,Khmel'nitskii71}, 
from analysis of the quantum spherical model\cite{Schneider76}
and from renormalization-group 
studies.\cite{Schmeltzer83,Sachdev97}  
We note that this $1/T^2$
behavior in the dielectric susceptibility of the quantum paraelectric 
has been observed
experimentally.\cite{Rytz80,Coleman06,Rowley07}
We summarize in Table I the link between the conventional
Casimir effect and finite-temperature behavior in the
vicinity of a QCP.

\section{Finite-Size Scaling in Time}

The spatial confinment of order parameter fluctuations near a classical 
critical point has been studied as a ``statistical mechanical Casimir effect'',
\cite{Krech94,Fisher78,Danchev00} and here we extend this treatment to
study the influence of temperature near a QCP using finite-size scaling (FSS)
in imaginary time.
This scaling approach is 
strictly valid in dimensions less than
the upper critical dimension.
Quantum critical ferroelectrics in $d=3$ lie at the marginal dimension
($D = d + z = 4$), 
so the scaling results are valid up to logarithmic corrections,
which we discuss later (Sec. VI); here $z=1$ refers to a linear
dispersion relation, $\omega = c_s q$.

\figwidth=5in
\fg{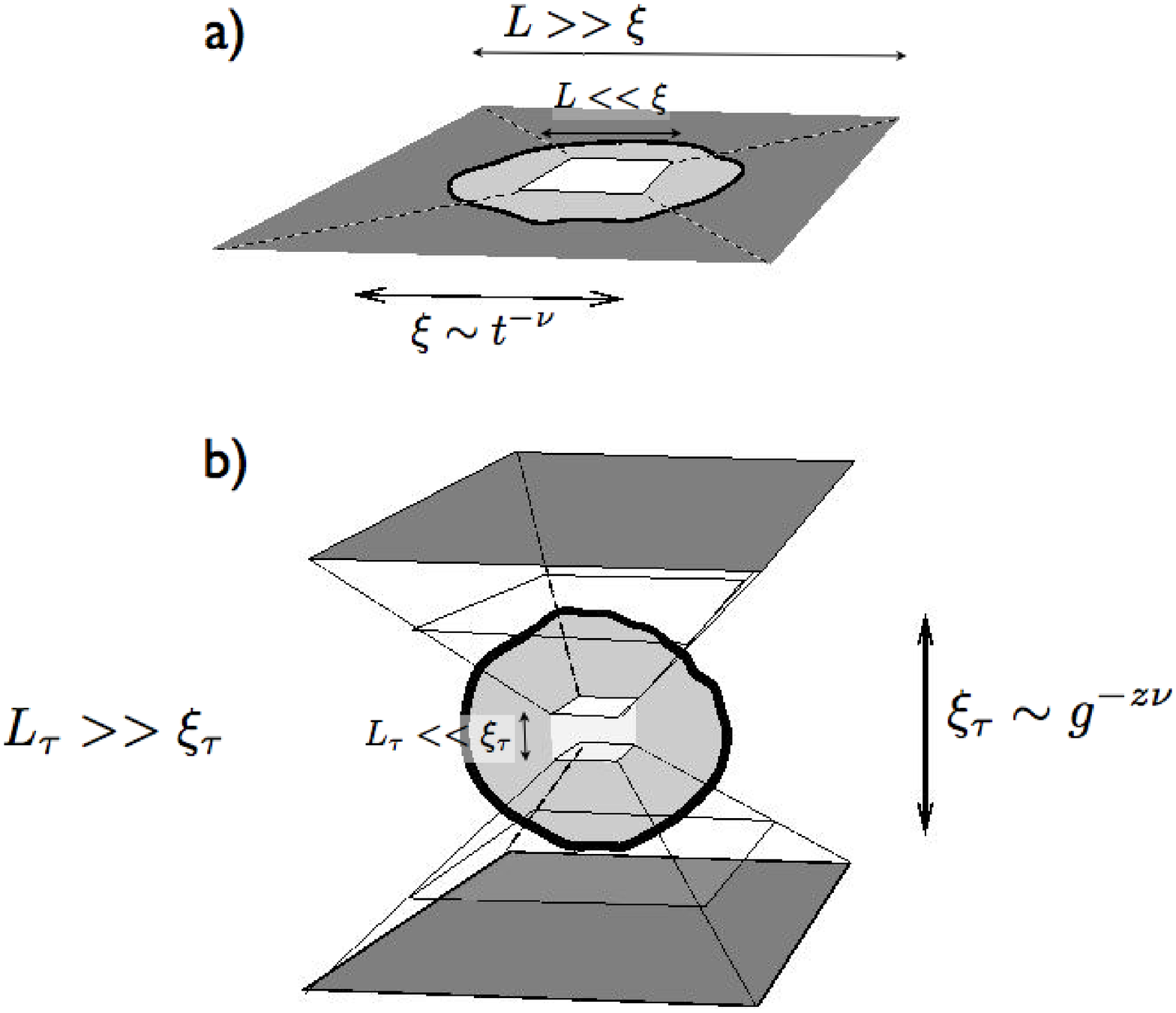}
{Schematic of finite-size effects (a) at a classical 
and at a (b) quantum critical point where the appropriate
lengths are defined in the text.}
{fig2}

Following the standard FSS procedure,\cite{Cardy96,Brezin85,Rudnick85}
we impose boundaries on the system near its critical point. For a classical 
system with tuning parameter $t= \frac{T-T_{c}}{T_{c}}$ and correlation 
length $\xi$, we confine it in a box of size $L$ and then
write the standard FSS scaling form
\begin{equation}
\chi = t^{-\gamma} f\left(\frac{L}{\xi} \right)
\label{cscaling}
\end{equation}
for the susceptibility.\cite{Cardy96,Brezin85,Rudnick85,Chamati00}  
Similar reasoning can be used when a system is near its QCP. Here
temperature is no longer a tuning parameter, this role is taken over
by an external tuning field $g$. Temperature now assumes a new role as 
a boundary condition in time.
Introducing 
a fixed $L_\tau$ (see Fig. 2b) associated with a finite $T$, while
replacing $t\rightarrow g$, the quantum critical version of  (\ref{cscaling})
is
\begin{equation}
\chi =  g^{-\gamma} {\Phi} \left(\frac{L_\tau}{\xi_{\tau}}\right)  
\label{gscaling}
\end{equation}
where $g$ is the tuning parameter.
The dispersion relation $\omega = c_s q^z$ yields $[\xi_\tau] = [\xi^z]$; this combined
with $\xi \sim g^{-\tilde\nu}$ leads to 
$\xi_{\tau} \sim g^{-z\tilde\nu}$.
We therefore write
\begin{equation}
\chi =  g^{-\gamma} \Phi \left(\frac{L_\tau}{g^{-z\tilde\nu}}\right)  
\label{chig}
\end{equation}
where $\Phi (x) \sim x^p$ is a crossover function where $p$ is determined
by the limiting values of $\Phi(x)$; 
when $x \rightarrow 0$, we expect $\chi = \chi(L_\tau)$,
whereas we should recover the zero-temperature result ($\chi \sim g^{-\gamma}$) 
when $x \rightarrow \infty$.  
Therefore we obtain
\begin{equation}
\chi \sim
g^{-\gamma} \left(\frac{L_\tau}{g^{-z\tilde{\nu}}}\right)^{\frac{\gamma}{z\tilde{\nu}}} 
\sim L_\tau^{\frac{\gamma}{z\tilde{\nu}}} \sim T^{-\frac{\gamma}{z\tilde{\nu}}} 
\label{chiqmscaling}
\end{equation}
and the temperature-dependence ($L_\tau \propto 1/T$) emerges naturally from FSS arguments.
Therefore a ($T=0$) quantum critical point can influence thermodynamic
properties of a system at finite $T$ just as a finite-size system displays
aspects of classical critical phenomena despite its spatial constraints.
A schematic overview of the finite-size scaling arguments
we have presented here is displayed in Figure 3.

\figwidth=5in
\fg{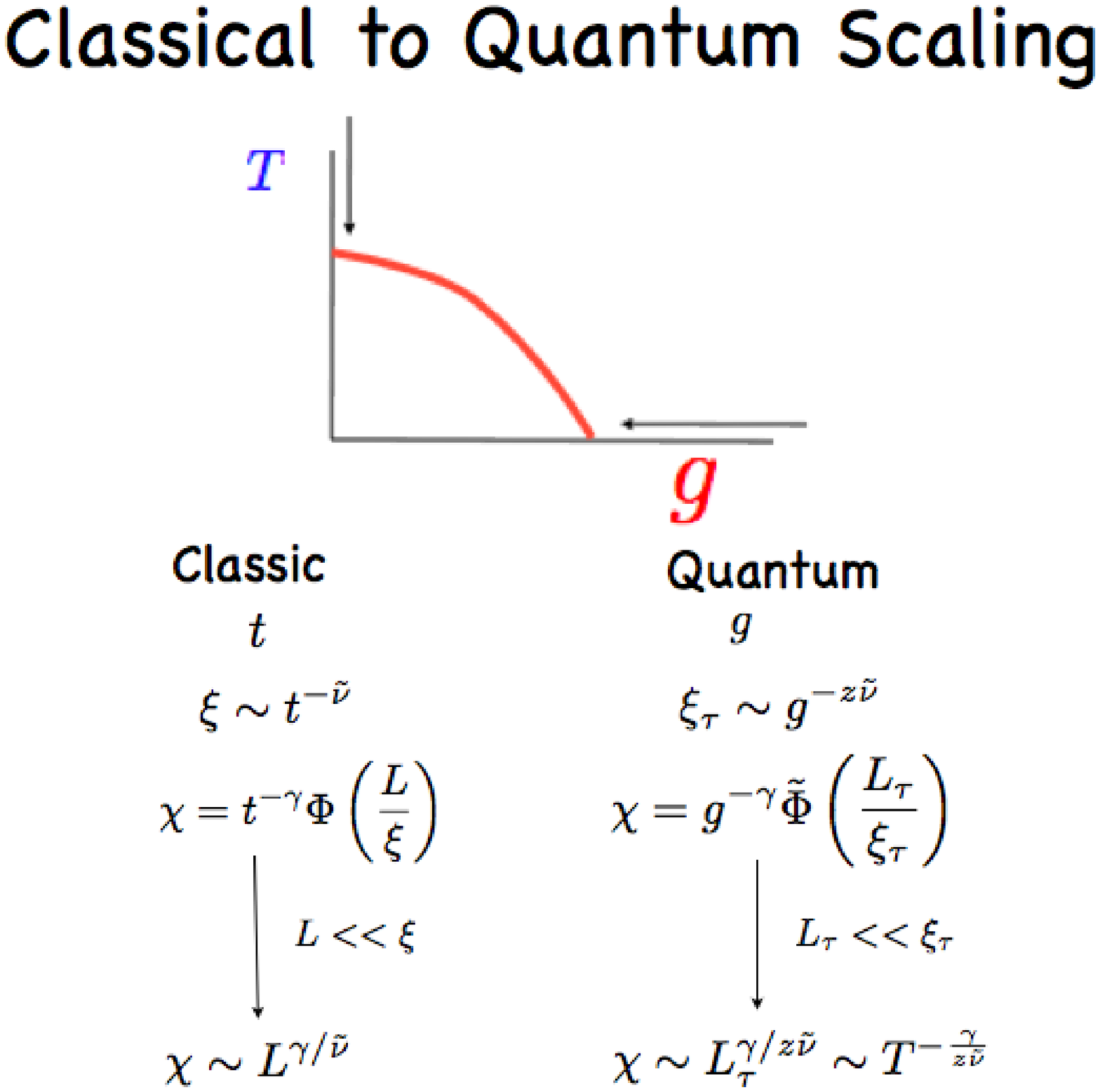}
{Overview of the finite-size scaling at classical and quantum critical points;
here $\tilde{\nu}$ is the exponent associated with the spatial correlation
length since $\nu$ has already been used in the text as a frequency.}
{fig3}  

The FSS approach can also yield the $T$-dependences
of the specific heat and the polarization of a quantum critical
paraelectric. At a finite temperature phase transition, to obtain the
specific heat capacity of a finite  size box with $L<<\xi$, we write
$f\sim t^{2-\alpha }F (L/\xi)\sim t^{2-\alpha } \left(
\frac{L}{t^{-\nu}} \right)
^{- ( 2-\alpha )/\nu}\sim L^{-(2-\alpha )/\nu}$. 
In a similar spirit, applying the quantum FSS analogies
($L\rightarrow
L_{\tau }, \ t\rightarrow g, \xi \rightarrow \xi_{\tau }^{z}=
g^{-z\bar \nu}$), we obtain
\begin{equation}
f_{qm}(L_\tau) \sim g^{2 -\alpha} \left(\frac{L_{\tau}}{g^{-z\tilde\nu}}\right)^{-\frac{(2 - \alpha)}{z\tilde\nu}} 
\sim L_\tau^{-\frac{2 - \alpha}{z\tilde\nu}} 
\sim T^{\frac{2-\alpha}{z\tilde\nu}}
\label{fqmT}
\end{equation}
so that the $T$-dependent specific heat is
\begin{equation}
c_v (T) = T\partial^2 f_{qm}/\partial T^2 
\sim T^{\frac{2-\alpha}{z\tilde{\nu}}-1}
\label{tcv}
\end{equation} 
in the vicinity of a QCP.  Similarly the $T$-dependence of the polarization is
$P(T) \sim T^{\frac{\beta}{z\tilde{\nu}}}$ 
and we note that
$P(E) = \partial f_{qm}/\partial E|_{g = 0} \sim E^{1/\delta}$ is 
$T$-independent, since finite-temperature
scaling does not affect field-behavior.

Simple scaling relations at classical and quantum criticality are summarized
in Figure 3. The key notion is that at a QCP, 
finite $T$ effects correspond to the limit $L_\tau << \xi_\tau$;
in this case $L_\tau$ becomes the effective correlation length in time, and
the $T$-dependences follow. We note that we expect this finite-size
approach to work for dimensions $d < d_c^u$ where there will be
logarithmic corrections to scaling in the upper critical dimension $d_c^u$.


Let us now be more specific with exponents for the quantum 
paraelectric case.
At criticality the observed $T$-dependence of the paraelectric susceptibility ($\chi$) 
can be found by a 
soft-mode analysis, \cite{Muller79, Lines77} and
therefore the exponents for the quantum paraelectric 
are those of the quantum spherical model.\cite{Schneider76}
For the case of interest ($D = d + z = 3 + 1 = 4$), the quantum spherical model
has exponents $\tilde{\nu} = 1/(D-2) = 1/2$ and $\gamma = 2/(D-2) = 1$, so that 
$\gamma_{th} = \frac{\gamma}{z\tilde{\nu}} = 2$ and
we recover the $\chi^{-1} \sim T^2$ scaling found earlier. 
Other specific $T$-dependences are displayed in Table II.
For $d=3$, we have $g \sim T^2$;
this relation was experimentaly observed~\cite{Venturini04,Wang01}  
in $STO$.
Finally we note that the FSS
that we have discussed suggests the 
``$\frac{\omega}{T}$'' scaling form
\begin{equation}
\chi(\omega,T) = \frac{1}{\omega^2} F \left(\frac{\omega}{T}\right)
\label{scalingform}
\end{equation}
that is similar to that observed in other systems at quantum
criticality;\cite{Aronson95,Schroeder00} this was previously
derived by more technical methods.\cite{Sachdev97}
Predictions for experiment are summarized in Table II.
We note that since we are in the upper critical dimension, there will
be logarithmic corrections to this scaling but we do not expect these
to be experimentally important for the temperature dependences described
here; however they will be considered later in the paper (Section VI).

\centerline{\bf Table II. Observables for a $D=4$ QPE in the Vicinity of a QFCP}
\begin{center}
\label{qscalingtab}
\begin{tabular}{|l||c|c|}
\hline
Observable & T-Dependences& g-Dependences
\\
& (g=0) & (T=0)\\
\hline
Polarization& 
$P \sim T^1$
&
$P \sim g^{\frac{1}{2}}$
\\
\hline
Susceptibility& 
$\chi \sim T^{-2}$&
$\chi \sim g^{-1}$
\\
\hline
\end{tabular}
\end{center}
\begin{center}
\begin{tabular}{|c|}
\hline
$P \sim E^{\frac{1}{3}}$\\
$\chi (\omega,T) = \frac{1}{\omega^2} F \left(\frac{\omega}{T} 
\right)$\\
\hline
\end{tabular}
\end{center}
\label{qscalingtab}

\section{Gaussian Theory: Illustration of Temperature as a Boundary Effect}
\label{MFT}

\subsection{The Gap Equation}\label{}

In this section we use the self-consistent 
Gaussian theory to illustrate how the $\chi(T)$
found via FSS in time appears from a more microscopic
approach; we also study the crossover behavior
between the classical and the quantum critical points.
This approach is equivalent to the self-consistent
one-loop approximation\cite{Moriya85} that is used in the 
context of metallic magnetism. 

The soft-mode treatment has been described extensively 
elsewhere;\cite{Muller79,Lines77,Moriya85} here we briefly outline
the derivation of the gap equation. 
The Lagrangian in Euclidean space-time, ${\cal L}_E$ in (\ref{SEuclidean}), for 
displacive ferroelectrics is
the $\phi^4$ model:
\begin{equation}\label{lagrangian}
{\cal L}_E \rightarrow 
\frac{1}{2}  \left[ (\partial_{\tau }{P})^{2}+ (\nabla
P)^{2}+
r P^{2} \right]  
 + \frac{\gamma_c}{4}  P^4.
\end{equation}
which determines the partition function.
Notice that in writing (\ref{lagrangian}), we have chosen rescaled units in which the
characteristic speed of the soft mode $c_{s}=1$. 
In 
a self-consistent
Hartree theory, 
interaction feedback is introduced via its renormalization of
quadratic terms; this procedure is equivalent 
to replacing ${\cal L}_{E}$
by the Gaussian Lagrangian
\begin{equation}\label{selfc}
{\cal L}_{G} = \frac{1}{2}  P
\left[ - \partial^{2}_{\tau }
- \nabla^{2} + r + \Sigma 
 \right]P 
\end{equation}
where 
\begin{equation}
\Sigma =3\gamma_c\langle P^{2}\rangle
\label{selfcon}
\end{equation}
is the Hartree self-energy (see Fig. 4).  
We note that this mode-mode coupling theory is exact for the ``spherical model'' generalization
of $\phi^4$ theory in which the order parameter has $N$ components and $N$ is taken to
infinity.  

The Green's function can now be determined from Dyson's equation, shown diagrammatically in Figure 4,
and takes the form
\begin{eqnarray}\label{thegreenf}
G (q)\equiv  G (\vec{q},i\nu_n) =  \left[ (i\nu_n)^{2}- q^{2}
-r- \Sigma \right]^{-1},
\end{eqnarray}
so the action is diagonalized in this basis. 
The poles of $G (\vec{q},\omega)$ determine the 
dispersion relation  $\omega_{q}$
for the displacive polarization modes 
\begin{equation}
\label{omega}
\omega^2_{q} = q^2 + \Delta^2
\end{equation}
where here we 
have introduced the
gap function 
\begin{equation}\label{gapfunction}
\Delta^2(r,T)= r + \Sigma (r,T).
\end{equation}
This quantity 
vanishes at both quantum and classical critical points
where there are scale-free (gapless) fluctuations.   At the quantum
critical point where $T_{c}=0$, $\Delta (r_{c},0)= r_{c}+ \Sigma (r_{c},T=0)$, 
so that we can eliminate $r_{c}=-\Sigma (r_{c},T=0)$, to obtain
\begin{equation}\label{tricky}
\Delta^2 (r,T) 
 = \Omega_{0}^{2}+[\Sigma (r,T)-\Sigma (r_{c},0)],
\end{equation}
where $\Omega_{0}^{2}=(r -r_{c})= g $.

\figwidth=5in
\fg{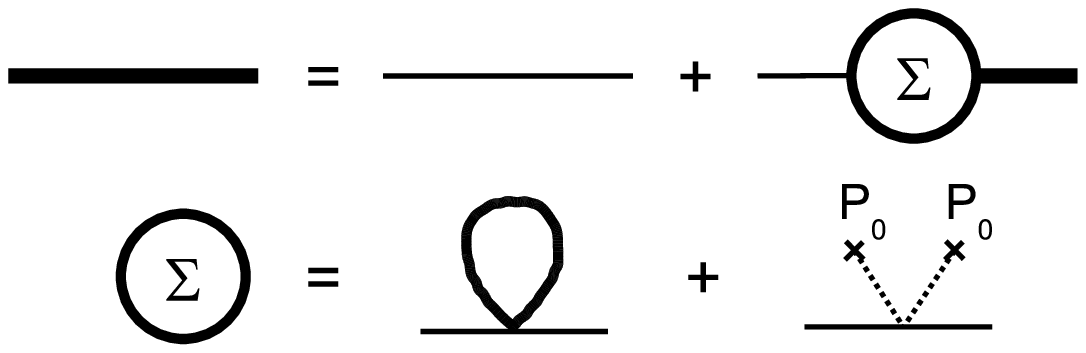}
{Diagrammatic Representation of (a) the Dyson Equation and (b)
the Gaussian Self-Energy 
where $P_0 = 0$ in the paraelectric state (and is finite in the ferrelectric phase.}
{fig4}

The amplitude of the
polarization fluctuations is given by 
\begin{equation}
\label{P2ave}
\langle P^2 \rangle = - G(0,0) = - \frac{1}{\beta V} \sum_{q} 
G(q) e^{iqx} \vert_{x=0},
\end{equation}
so 
the self-consistency (\ref{selfcon}) condition $\Sigma = 3 \gamma_c
\langle P^{2}\rangle  $ can now be written 
\begin{equation}
\label{sigmagen}
\Sigma (r,T) = 
(-3\gamma_c) T \sum_n \int \frac{d^d{q}}{(2\pi)^d} G(q,i\nu_n),
\end{equation}
where $\Sigma (r,T)$ is the temperature-dependent self-energy. 
By converting the discrete Matsubara summation to a contour integral,
deformed around the poles $z= \pm \omega (q)$ in the dispersion
relation, we can convert this expression to the form
\begin{equation}\label{}
\Sigma (r,T) = 
3 \gamma_c
\int\frac{d^{d}q}{(2\pi)^{d}}
\frac{\left[ n_B({\omega_{q}}) + {\textstyle\frac{1}{2} }\right]}{\omega_{q}}
\end{equation}
where we denote $n_{B} (\omega)\equiv n_{B} (\omega,\beta )=  1/ (e^{\beta \omega}-1)$.
At the quantum critical point ($r = r_c$ and $T=0$), 
we have $\omega_{q}= q$ and $n (\omega_{q})=0$ so that 
\begin{equation}
\Sigma (r_{c},0) = 
3 \gamma_c
\int\frac{d^{d}q}{(2\pi)^{d}}
\frac{1}{2q} ,
\end{equation}
and using (\ref{tricky}), we can write the gap function as
\begin{eqnarray}\label{gapfunc}
\Delta^{2} &=&\Omega_{0}^{2}
+
3 \gamma_c
\int\frac{d^{d}q}{(2\pi)^{d}}
\left(\frac{\left[ n_B({\omega_{q}}) + {\textstyle\frac{1}{2}
}\right]}{\omega_{q}} -\frac{1}{2q} \right),\cr
\omega_{q}&=& \sqrt{q^{2}+\Delta^{2}}.
\end{eqnarray}

\subsection{$T$-Dependence of the Gap at the QCP.}\label{}

In the paraelectric phases, we can use the temperature-dependent gap
to determine the dielectric susceptibility $\chi$. Writing 
\begin{equation}
\chi  = \chi (q,\omega) \Biggr|_{\vec{q}, \omega=0}
= \langle P(q) P(-q) \rangle \Biggr|_{q=0} 
= - G (\vec{q},\omega) \Biggr|_{\vec{q}, \omega=0},
\label{dsuscep}
\end{equation}
we use (\ref{thegreenf}) and (\ref{gapfunction}) to express it as 
\begin{equation}
\label{chiinv}
\chi^{-1} (r,T) = \Delta^2 (r,T).
\end{equation}

At the quantum critical point
$\Omega_{0}^{2}=0$, so  the gap equation (\ref{gapfunc}) becomes
\begin{equation}
\label{sigmacrit}
\Delta^{2} (r_{c},T)= 3\gamma_c
\int_0^{q<q_{max}} \frac{d^{d}q}{(2\pi)^{d}}\Big\{ 
\frac{ \left[ n_B(\omega_{q}) + \frac{1}{2} \right]}{\sqrt{q^2 + \Delta^2}} 
- \frac{1}{2q} \Big\},
\end{equation}
where we have inputted the dispersion relation, (\ref{omega}), for $\omega_q$
in (\ref{sigmacrit}).
We notice that both thermal and quantum fluctuations contribute to this 
expression. 

Even though the mean field gap equation is only formally exact in the
spherical mean-field limit, it is sufficient to illustrate the
qualitative influence of $T$  on the gap at the QCP. In order to explore 
the cutoff-dependence of (\ref{sigmacrit}), we note
that in the ultraviolet limit of
interest, the last two terms 
can be expressed as
\begin{equation}
\frac{1}{2} \left\{ \frac{1}{\omega_q} - \frac{1}{q} \right\} 
= -\frac{\Delta^2}{4q^3},
\label{uvstuff}
\end{equation}
where there is complete cancellation when $\Delta = 0$ exactly at the QCP.  However just
slightly away from it, when $\Delta$ is finite, (\ref{uvstuff})
leads to a $q^{d - 3}_{max}$ scaling-dependence
of the integral in (\ref{sigmacrit}); therefore the cutoff is required
to ensure that (\ref{sigmacrit}) is finite in dimensions
$d>3$. 
However, in dimensions $d<3$, this integral is convergent in the
ultraviolet and the upper cutoff in (\ref{sigmacrit}) can
be entirely removed. Thus, for $d<3$, the only scale in the problem is
temperature itself. The integral is also convergent in the infrared provided
$d>1$.  The spatial dimensions $d=1$ and $d=3$ correspond to 
spacetime dimensions $D=d+1= 2$ and $D=d+1=4$, which are the
well-known lower and upper critical dimensions of the $\phi^{4}$
theory. This provides us with a dimensional window  $1<d<3$ where 
inverse temperature acts as a cut-off in time. 
In this range, the temperature-dependence of the gap 
\begin{equation}
\Delta (T) = \alpha_{d}  T
\label{gapt}
\end{equation}
is independent of the strength of the coupling constant $\gamma_c$ and
the cutoff, a feature that can be illustrated already within
mode-coupling theory. Recalling that 
$\Delta(T) = \frac{\alpha}{L_\tau}$
and $\alpha \equiv \alpha_d$
(see (\ref{definegap}) and Fig.1), 
we note that
confirmation of (\ref{gapt}) 
is consistent with our earlier discussion (see after (\ref{ltau2}))
that $\xi_\tau$ is independent of coupling constant; in this dimensional window,
temperature is a boundary effect in (imaginary) time
and is the only temporal scale in the problem.  

\figwidth=5in
\fg{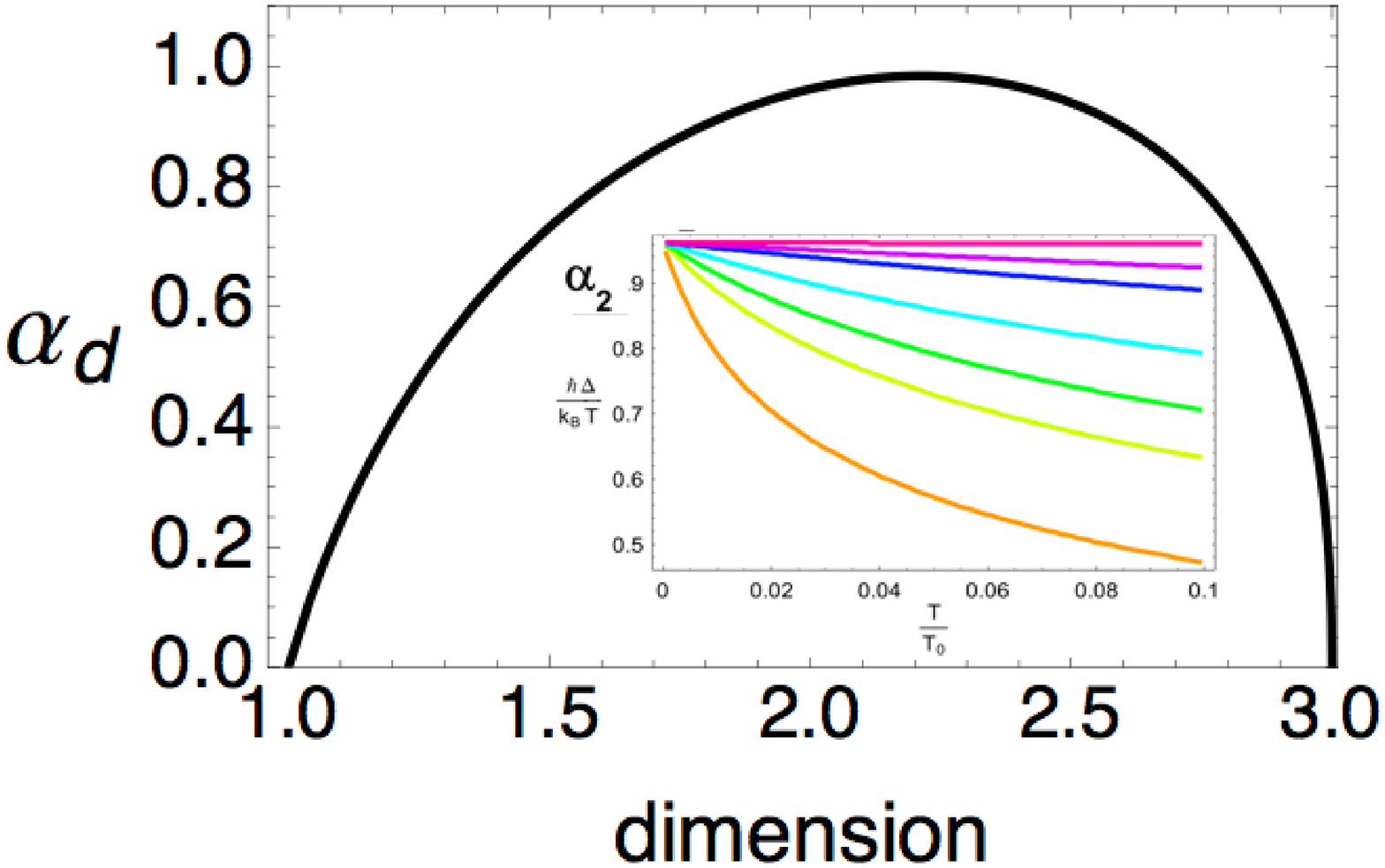}
{Dependence of $\alpha_d(T \rightarrow 0)$ on dimensionality $d$.
Inset:  $T$-dependence of $\Delta/T$
for $d=2$ and couplings in the range 
$0.01 < \frac{\gamma_c (d=2)}{q_{max}} <  5.0$; here $T_0$ is the temperature scale where $\xi \sim a$ and we note that $\lim_{T\rightarrow 0} \alpha_2$ is 
independent of $\gamma_c$.}  
{fig5}

In order to calculate $\alpha_{d}$, we rewrite the gap equation at criticality
as
\begin{equation}
\frac{\Delta^{2}}{T^{2}}= \alpha^{2}= 
\frac{3\gamma_c}{T^2} \Gamma_{d}
\int_0^{\infty } \frac{q^{d-1}dq}{(2\pi)^{d}}\Big\{ 
\frac{ \left[ n_B(\omega_{q}) + \frac{1}{2} \right]}{\omega_{q}} 
- \frac{1}{2q} \Big\},
\end{equation}
where $\Gamma_{d}q^{d-1}dq $ $(\Gamma_{d}= \frac{2 \pi^{d/2}}{\Gamma (d/2)})$ is the d-dimensional volume measure.
Rescaling $\Delta = \alpha_d T$  and $q= u T$, we obtain
\begin{eqnarray}\label{lungo}
F_{d}[\alpha ] =  T^{3-d}\alpha_d^{2}/\gamma_c
\end{eqnarray}
where
\begin{equation}\label{lungo2}
F_{d}[\alpha ]=\frac{3}{(2\sqrt{\pi})^{d} \Gamma(d/2)}
\int_0^{\infty } {u^{d-1}du}\Big\{ 
\frac{\coth (\frac{1}{2}\sqrt{u^{2}+\alpha^{2}}) }{\sqrt{u^{2}+\alpha^{2}}} 
- \frac{1}{u} \Big\}.
\end{equation}
For $d<3$,  the temperature prefactor  on the right-hand side of
(\ref{lungo}) vanishes 
$T\rightarrow 0$, so a consistent solution requires $\alpha_{d}$ to satisfy
\begin{equation}\label{}
F_{d}[\alpha_{d}]= 0.
\end{equation}
At a small finite temperature, we can expand around
$\alpha=\alpha_{d}+ \delta \alpha (T)$, to obtain 
\begin{equation}\label{deltagamma}
\Delta (T)= \alpha_{d}T + \left(\frac{\alpha_{d}^{2}}{\gamma_c F'[\alpha_{d}]} \right)T^{4-d}.
\end{equation}
Thus in dimensions $d<3$, the dominant low temperature behavior is independent
of $\gamma_c$, the strength of the mode-mode coupling, which enters into the
subleading temperature dependence. 

The necessity of separating out the singular part of equation (41)
was pointed out to us by Chamati
and Tonchev;\cite{Chamati09} (41) was incorrectly treated in
an earlier version of this paper.  Following their approach, we can
rewrite (41)
as 
\begin{equation}\label{lungoa}
F_{d}[\alpha ]=\frac{3}{(2\sqrt{\pi})^{d} \Gamma(d/2)}\biggl[
\int_0^{\infty } {u^{d-1}du}\left(
\frac{\coth (\frac{1}{2}\sqrt{u^{2}+\alpha^{2}}) -1}{\sqrt{u^{2}+\alpha^{2}}} 
\right)
+
\overbrace {\int_0^{\infty } {u^{d-1}du}
\left(\frac{1}{\sqrt{u^{2}+\alpha^{2}}}-\frac{1}{u} \right)
}^{
\frac{\alpha^{d-1}}{2\sqrt{\pi}}
\Gamma (\frac{d}{2})\Gamma (\frac{1-d}{2})
}\biggr],
\end{equation}
yielding
\begin{equation}\label{lungob}
F_{d}[\alpha ]=\frac{3}{(2\sqrt{\pi})^{d} \Gamma(d/2)}
\int_0^{\infty } {u^{d-1}du}\Big\{ 
\frac{\coth (\frac{1}{2}\sqrt{u^{2}+\alpha^{2}}) -1}{\sqrt{u^{2}+\alpha^{2}}} 
\Big\} -\frac{12}{(2\sqrt{\pi})^{d+1}}\alpha^{d-1}
\left(\frac{\Gamma (\frac{5-d}{2})}{(d-1) (3-d)} \right),
\end{equation}
provided $(1\leq d \leq 3)$.
The first term in this expression is  a smooth positive 
function of $d$ and $\alpha $, whereas the second is a singular
negative function of $d$ with poles at $d=1$ and $d=3$.
The numerical solution of $F_{d}[\alpha_{d}]=0$ can then be determined and
is presented in Fig. \ref{fig5}.
We note that this result indicates that
$\alpha_{d}$ vanishes in the vicinity of $d\sim 3$ as $\alpha_{d}\sim
\sqrt{3-d}$, consistent with previous $\epsilon$ calculations.\cite{Sachdev97}


In Figure 5 we display the dependence of $\alpha_d$ on dimensionality $1<d<3$.
The temperature-dependence of the gap in two dimensions 
is shown in the inset of Fig. 5, where we see that 
$ \lim_{T \rightarrow 0}\alpha_2 \equiv 0.96$
is the same for all couplings.
According to (\ref{lungo}) and (\ref{lungo2}), we write
$\alpha_3^{2} = \gamma_c F_{3}[\alpha]$ and solve for $\alpha_3$ in the limit of
upper cutoff $u_{max} = q_{max}/T \equiv 2\pi T_0/T >> \alpha_3$,
\begin{equation}
\label{alpha3}
\alpha_3 (T,\gamma_c) \sim \sqrt{\frac{\gamma_c}
{1 + \gamma_c (\frac{3}{8\pi^2}) \ln (\frac{4\pi T_0}{T})}},
\end{equation}
where again we do not consider logarithmic corrections to $\alpha_3$.
In the limit of strong coupling, 
$\alpha_3 \sim \big[ \ln (\frac{4\pi T_0}{T}) \big]^{-1/2}$
is $\gamma_c$ independent.
For weak coupling, the situation relevant here,
$\alpha_3$ is indeed a function of $\gamma_c$
but remains independent of temperature so that 
$\Delta \sim T$ according to (\ref{gapt}); temperature-dependences
derived here should therefore be in agreement with those found
from a scaling perspective whenever direct comparison is possible.

\subsection{Temperature-Dependent Dielectric Susceptibility }\label{}

To provide an explicit illustration  of the above calculations, 
we now use 
(\ref{gapfunc}), and (\ref{chiinv}) to numerically determine
the temperature-dependent paraelectric susceptibility in the approach to the  
quantum critical point (QCP) in $d=3$. 
We obtain $\chi^{-1}(T) = \Delta^{2} \sim T^2$ 
for the approach $r =r_c$ in agreement with previous results and discussion.  We note that a similar analysis
in the vicinity of the classical phase transition leads to the expected
Curie susceptibility ($\chi^{-1} (T\rightarrow T_c^+ >> 0) \sim T$) since in this
(high) temperature regime the Bose function in (\ref{sigmacrit}) scales as $\frac{T}{\omega}$.  We also remark that
if we assume that $\omega \equiv \tilde{\omega}_0$ with no q-dependence
then we recover the 
Barrett\cite{Barrett52} expression 
$\chi^{-1} \sim A + B \coth \frac{\hbar \tilde{\omega}_0}{k_BT}$;
because the disperson is constant and q-independent this approach is not applicable
near quantum criticality where the gap vanishes and the q-dependence becomes important.

One more point needs to be considered before we proceed with our self-consistent Hartree theory. 
In the self-consistent Hartree theory (SCHT) of the 
ferro-electric phase,  the polarization field $P_{0}$
acquires a nonzero value.
$P_0$ enters the Lagrangian ${\cal{L}}_E$ in (\ref{lagrangian})
as $P = P_0 + \delta P$,
where $\delta P$ are fluctuations of the polarization field around its
mean value, $P_0$ ($P_0 = 0$ in the
paraelectric phase).
The self energy (\ref{selfcon}) becomes
\begin{equation}
\label{Pequilibrium}
\Sigma = 3\gamma_c \langle P^2 \rangle = 3\gamma_c \, \big( P_0^2 + \langle \delta P^2 \rangle \big)
\end{equation}
as indicated diagrammatically in Figure 4.
The equilibrium value $P_0$ is easily obtained by introducing an
electric field into the Lagrangian by replacing 
${\cal L}_E \to {\cal L}_E + E \cdot P$, then seeking the stationary
point $\delta S/\delta P_{0}=0$ which gives
$\langle rP_0 + 3\gamma_c \delta P^2 P_0 + \gamma_c P_0^3 - E \rangle
= 0$,
or
\begin{equation}
\label{P0}
r + \Sigma - 2 \gamma_c P_0^2 = \frac{E}{P_0} = 0 
\end{equation}
%
at zero electric field. According to (\ref{gapfunction}), 
$\Delta^2 (r,T) = r + \Sigma(r,T)$, so that
the spectral gap in the ferroelectric phase is
\begin{equation}
\label{deltaferro}
\Delta_{f}^2 (r,T) = 2\gamma_c P_0^2 (r,T) > 0.
\end{equation}

\figwidth=3in
\fg{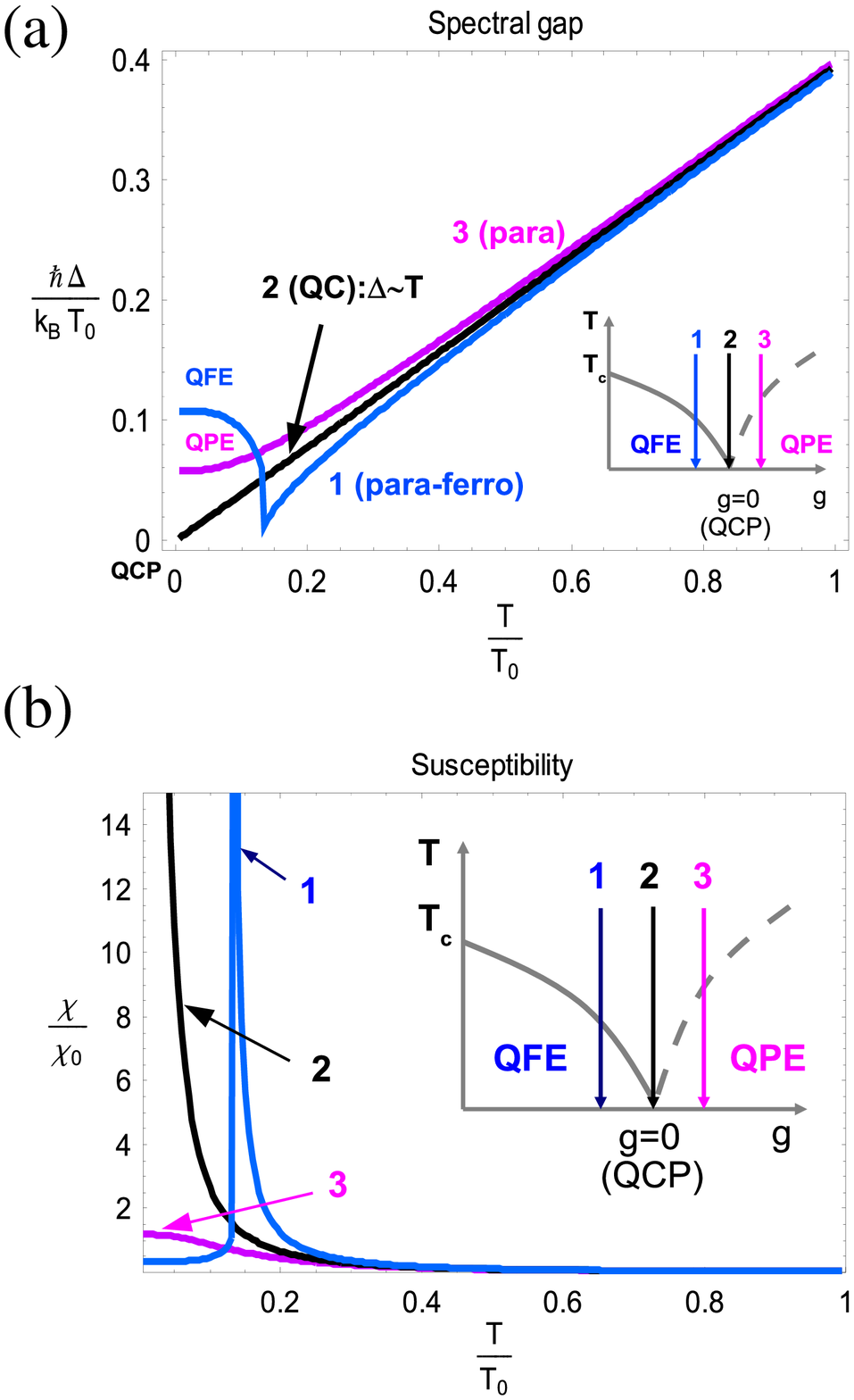}
{Temperature-dependence of the a) spectral gap and b) the 
dielectric susceptibility
for three temperature scans defined in the schematic
inset; here $g = r - r_c$.}
{fig9}

In Fig. \ref{fig9} a) we plot
the calculated temperature-dependent spectral gap $\Delta(r,T)$ for three
different values of $r$ 
as indicated in its schematic inset.
As expected, for (2) the spectral gap closes
exactly at $T=0$ leading to a linear dispersion relation, $\omega = q$ at the 
QCP. We note
that in the quantum paraelectric (QPE), 
$\Delta$ (or $\chi^{-1}$) is
constant.  In the quantum ferroelectric (QFE) again $\Delta$ is
constant; though there exists a classical paraelectric-ferroelectric transition at $T=T_c$ where
$\chi^{-1} \sim (T - T_c)$.  The static dielectric susceptibility in the vicinity of the QCP
(low T) is presented in the same three $r$ regimes in Fig.\ref{fig9} b) where we see that 
in the QPE regime 
$\chi (T\rightarrow 0)$ saturates, at the QCP $\chi(T) \sim T^{-2}$ and diverges as $T \rightarrow 0$.
In the QFE, the susceptibility also saturates at low temperatures, though the Curie law is
recovered in the vicinity of the classical transition at $T=T_c$.
\figwidth=5in
\fg{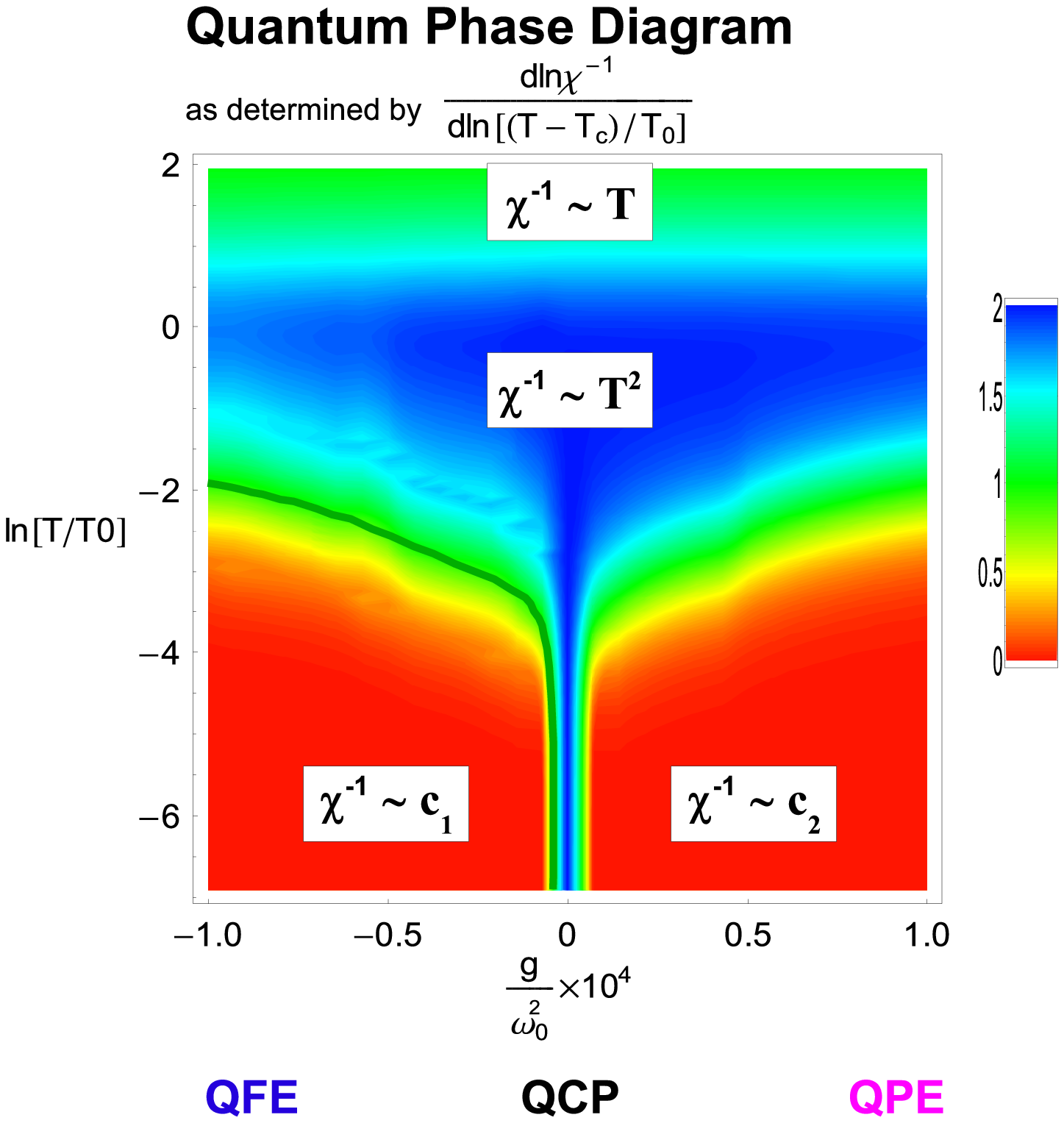}
{T-g Phase Diagram as determined by a self-consistent analysis of the 
dielectric susceptibilty. The power law exponents are depicted in different 
colors via the function
$\frac{d ln \chi^{-1}}{d ln (T-T_c)/T_0}$. This expression selects the
exponent $2$ (blue region) for $\chi^{-1} \sim T^2$ ($T_c \equiv 0$ for QCP),
exponent $1$ (green region) for classical Curie behavior $\chi^{-1} \sim (T-T_c)$ and exponent
$0$ (red region) for a constant susceptibility. 
}
{fig11}

Figure \ref{fig11} shows the phase diagram that 
results from the self-consistent Hartree theory.
This figure serves to emphasize 
how the strictly zero temperature QCP 
gives rise to a quadratic power law dependence of the inverse
susceptibility on temperature over 
a substantial region of the $T-g$ phase diagram. 

The crossover temperature, $T_0$, between Curie ($\chi^{-1} \sim T$) and 
Quantum Critical ($\chi^{-1} \sim T^2$) behavior in the susceptibility 
is defined by the expression
\begin{equation}
\label{T0equation}
T_0 \approx \frac{\hbar \omega_0 }{2\pi k_B},
\end{equation}
where $\omega_0 = \frac{c_s}{a}$ is the characteristic soft mode frequency, $c_s$ is the
soft-mode velocity and $a$ is the lattice spacing.
Here we have assumed a simple bandstructure $\omega(q) = \omega_0 \sin qa$
such that $c_s = 
\left . \phantom{\int}\frac{d\omega}{d q} \right\vert_{q=0} = 
\left . \phantom{\int} ( \omega_0 a)\cos qa \right\vert_{q=0} 
$ so
so that $\omega_0 = \frac{c_s}{a}$ as stated above. 
The factor of $2 \pi$ in the denominator of (\ref{T0equation}) 
results from the observation that the 
separation of the poles
of the Bose and Fermi functions in the complex frequency plane is 
$\Delta \nu_n =  2 \pi  k_{B}T$, which sets 
the natural conversion factor between 
temperature and frequency to be $2 \pi k_{B}$.  $T_0$
also corresponds to the temperature when the correlation length is
comparable to the lattice constant ($\xi \sim a$); here the correlation length
$\xi \equiv \frac{c_s}{\Omega_0} \sim g^{-1/2}$ (see (\ref{chig})).
Neutron scattering measurements\cite{Yamada69}
of the dispersion relation
indicate that the soft mode velocity in STO is $c_s \approx 10^4 m/s$
and the lattice constant has been measured~\cite{Haeni04} to be
$a_{STO} = 3.9 \times 10^{-10} m$; therefore
$T_0 \approx 30 K$.
We note that with $O^{18}$
substitution, the ambient pressure Curie temperature\cite{Venturini04} is 
$T_c \sim 25 K$. 
Using the values of $c_s$ and $a_{STO}$
from above, we get
$\omega_0 \approx 2.6 \times 10^{13} Hz$ in $SrTiO_3$.
The typical frequency $\Omega_0 = g^{1/2}$ (spectral gap at zero temperature) 
at which one observes the change of behavior in the dielectric susceptibility 
(blue region) is thus from Figure 7, 
$\Omega_0 \approx 10^{-2} \: \omega_0 = 2.6 \times 10^{11} Hz$.
Indeed, Raman scattering on ferroelectric $SrTi ^{18}O_3$ ($T_c = 25 K$)
shows that the zero temperature Raman shift~\cite{Takesada06}
is about $10 \: cm^{-1}$ 
which translates into $\Omega_0 \approx 3 \times 10^{11} Hz$,
in good agreement with our calculation.

\section{Coupling to Long-Wavelength Acoustic Modes}

\subsection{Overview}

In a conventional solid, broken translational symmetry leads to three
acoustic Goldstone modes. At a ferroelectric QCP,
these three modes are supplemented by one or more optical zero
modes.  This coexistence of acoustic and optic zero modes is a unique
property of the ferroelectric QCP, and in this section we examine how their 
interaction influences observable properties.

The  gap of the optical modes in a ferroelectric is sensitive to the
dimensions of the unit cell and couples linearly to  the strain field.
This leads to an inevitable coupling between the critical optical
mode and the long-wavelength acoustic phonons that must be considered.
To address this issue, we 
consider the effect of a coupling $\eta $ between the soft polarization 
and the strain field created by 
a single long-wavelength acoustic phonon mode.
Softening of the polar transverse optic (TO) mode
near the QCP enhances the effect of this coupling. 
Using dimensional analysis  we find that the coupling  between the TO and LA
mode is marginally relevant in the physically important dimension
$d=3$, and thus can not be ignored. The main result of the
analysis is that the acoustic phonons act to soften and reduce
the quartic interaction between the optic phonons. Beyond a certain
threshold $\eta>\eta_c$, this interaction becomes attractive, 
leading to the development
of a reentrant paraelectric phase at finite temperatures.
We note that such a coupling
to acoustic phonons has been considered previously,\cite{Khmel'nitskii71}
and here we are rederiving and extending prior results in a contemporary
framework.

\subsection{Lagrangian and Dimensional Analysis}

We introduce the coupling of the polarization ($P (\vec x,\tau)$) and the acoustic phonon
($\phi (\vec x,\tau)$) fields as a coupling of the polarization to strain
$- \eta \nabla \phi P^2$; we then write the Lagrangian~\cite{Khmel'nitskii71} as
\begin{equation}
\label{lagrangianac}
{\cal L}_E [P, \phi] = {\cal L}_E [P] + \frac{1}{2}\big[ (\partial_\tau \phi)^2 + \tilde{c}^2 (\nabla \phi)^2 \big] 
- \eta \nabla \phi P^2,
\end{equation}
where ${\cal L}_E [P]$ is our previous Lagrangian without acoustic coupling given in (\ref{lagrangian}).    
Here the constant $\eta$ is the coupling strength to the acoustic phonon; the 
latter's  
dynamics are introduced in the bracketed terms  of
(\ref{lagrangianac}). 
Since we are using units in which the velocity of the soft optical
phonon is one, $\tilde{c}= \frac{c_{a}}{c_{s}}$ 
is the ratio of the acoustic to the soft optical phonon velocities.
We will discuss the  restoration of dimensional constants 
in  (\ref{lagrangianac}) when we make comparison to experiment in
Section V. F.

We begin with a dimensional analysis of the couplings to assess their relative importance in
the physically important dimension $d = 3$.
In order to do so, we
introduce the renormalization group (RG) flow by rescaling length, time, momentum 
and frequency
\begin{equation}
x' = \frac{x}{\Lambda}, \quad \tau' = \frac{\tau}{\Lambda}, \quad q' = q \Lambda, \quad \nu' = \nu \Lambda,
\end{equation}
with constant $\Lambda > 1$ 
representing flow away from the infrared (IR) limit of the QCP, that is flow from small to large momentum and
frequency.
In terms of the rescaled variables $x'$ and $\tau'$, the action (\ref{SEuclidean})
with Lagrangian (\ref{lagrangianac}) in $d+1$ dimensions 
becomes
\begin{eqnarray}
\label{actionrescale}
S[P,\phi] &=& 
\int_0^\beta d\tau \int d^d x {\cal L}_E [P, \phi] \cr
&=&
\int_0^{\beta/\Lambda} d\tau' \int d^d x' \Lambda^{d+1} \Big\{ 
\frac{1}{2} \Lambda^{-2} \Big [
(\partial_{\tau'} P)^2 + (\nabla' P)^2 + (\partial_{\tau'} \phi)^2 + (\nabla' \tilde{c} \phi)^2 \Big ] \cr
&+& \frac{1}{2} \Omega_0^2 P^2 
 + \frac{1}{4} \gamma_c P^4 - \eta \Lambda^{-1} \nabla'\phi P^2 
\Big\}.
\end{eqnarray}
We emphasize that we write $\Omega_0^2 = r - r_c$ as the coefficient of the $P^2$ term 
in the Lagrangian $L_E[P]$ (\ref{lagrangian}), entering (\ref{lagrangianac}) in (\ref{actionrescale}),
since our RG flow starts from the QCP ($r = r_c$).
Rescaling 
$P$, $\phi$, 
$\Omega_0^2$, $\gamma_c$ and $\eta$,
so that the action (\ref{actionrescale}) assumes its initial form, 
we write
\begin{equation}
\label{rescalingrel}
P' = P \Lambda^{\frac{d-1}{2}}, \quad \phi' = \phi \Lambda^{\frac{d-1}{2}},
\quad (\Omega_0^2)' = \Omega_0^2 \Lambda^2, \quad \gamma_c' = \gamma_c \Lambda^{3-d}, \quad \eta' = \eta \Lambda^{2 - \frac{d+1}{2}},
\end{equation}
which leads to
\begin{equation}
\label{actionprime}
S[P,\phi] = \int_0^{\beta/\Lambda} d\tau' \int d^d x' {\cal L}_E [P', \phi'].
\end{equation}
Now the fields, the mass term and the coupling constants flow to new values leaving the action unperturbed.
We remark that the upper cuttoff in the imaginary time dimension is replaced by infinity as 
the temperature $T \sim \frac{1}{\beta}$ approaches zero. 

Analyzing the RG expressions in (\ref{rescalingrel}), we find that the $\Omega_0^2$ term
grows as we flow away from the QCP IR limit; therefore it is a relevant
perturbation parameter independent of dimension $d$.
This is consistent with the fact that 
$\Omega_0^2 = r - r_c = g$ {\sl tunes} the
system away from the QCP.
Similarly we find that couplings $\gamma_c$ and $\eta$ grow (relevant) in 
dimension $d<3$, decrease (irrelevant) in dimension $d>3$, and 
don't change (marginally relevant)
in $d=3$.
We see that in this case ($d=3$) the coupling to acoustic phonons ($\eta'$) is equally important
as the mode-mode coupling ($\gamma_c'$) and thus has to be included to 
the Gaussian model.

Let us now briefly summarize what we know about $\gamma_c$ before we proceed to the 
discussion of the acoustic coupling $\eta$.
In section IV B we found that the spectral gap $\Delta$ is independent
of $\gamma_c$ for dimensions $1<d<3$ in the zero temperature limit (see Fig. 5).
This is in agreement with the above results, where $\gamma_c$ is 
a relevant perturbative parameter;
more precise analysis~\cite{Cardy96} shows $\gamma_c$ flowing to the nontrivial 
Wilson-Fisher fixed point $\gamma_c^*$. Here
all the system properties become functions of 
$\gamma_c^* + \delta \gamma_c \approx \gamma_c^*$, and so are $\gamma_c$-independent.
On the other hand, in dimensions $d>3$ and $d=3$, $\gamma_c$ 
flows to zero (with logarithmic corrections in the marginal case).
In these cases the system properties are functions of $\delta \gamma_c$
 and thus are $\gamma_c$-dependent; we have already seen an example of this behavior
in the specific case of the $d=3$ spectral gap.

\subsection{Gap Equation}

\figwidth=3in
\fg{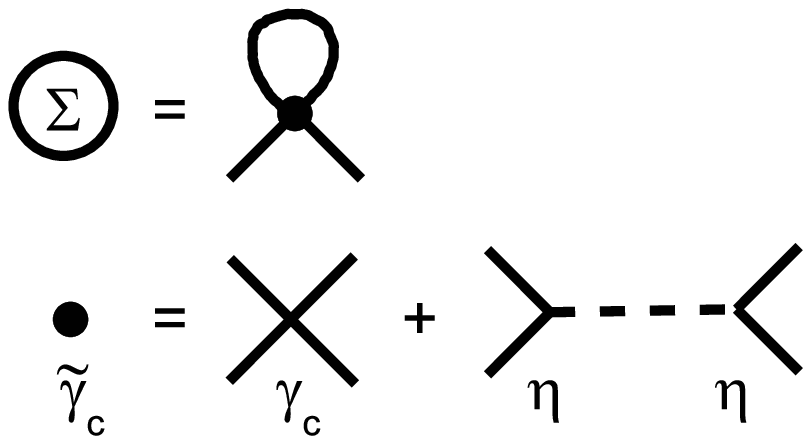}
{Diagrammatic representation of the self-energy that includes
coupling to both optical and acoustic phonons. Here $\tilde{\gamma_c}$ is
the renormalized coupling, including the exchange of an acoustic phonon.}
{fig12}

We are now ready to explore how the system's low-temperature behavior changes in the presence
of acoustic phonons in dimension $d=3$.
Let us look first at the LA phonon field $\phi$.
Following the procedure of Section IV A, we find the acoustic Green's function and dispersion 
relation from (\ref{lagrangianac}) to be 
\begin{equation}
\label{greensfac}
D(q) \equiv D (\vec q, i\nu_n) = [(i\nu_n)^2 - \tilde{c}^2 q^2]^{-1},
\end{equation}
\begin{equation}
\label{omegaac}
\omega_a(q) = \tilde{c} q.
\end{equation}

We emphasize the $P^2$-dependency of the new interaction term, $-\eta \nabla \phi P^2$, in  
the Lagrangian (\ref{lagrangianac}).
Therefore it contributes to the polarization self-energy as an additional term inside the brackets
of (\ref{selfc}).
This new contribution arises due to nonzero second-order perturbation and is schematically
shown in Figure 8, where the solid line represents the soft polarization TO Green's function 
(\ref{thegreenf}) and the dashed line represents the LA Green's function (\ref{greensfac}).
We note that the interaction represented by a {\it dot} in the self-energy consists of
a contribution each from the coupling $\gamma_c$ and $\eta$.
Thus we can write the polarization self-energy $\Sigma$ as a sum of these two terms 
\begin{eqnarray}
\label{sigmaacgen}
\Sigma (r,T) &=& \Sigma_{\gamma_c} (r,T) + \Sigma_\eta (r,T) \cr
             &=& 
(-3\gamma_c) T \sum_n \int \frac{d^3{q}}{(2\pi)^3} G(q,i\nu_n) \cr
             &+& 
4\eta^2 T \sum_n \int \frac{d^3{q}}{(2\pi)^3} q^2 G(q,i\nu_n) D(q,i\nu_n),
\end{eqnarray}
where $\Sigma_{\gamma_c}$ is the Hartree self-energy (\ref{sigmagen}) previously calculated in Section IV A.
We remark that the $q^2$ term in the integral for $\Sigma_\eta$ arises due to form of the interaction
($\nabla \phi$).
Converting the Matsubara summation to a contour integral, deformed around the poles 
$z_p = \pm \omega_p(q)$
and $z_{a} = \pm \omega_{a}(q)$ in the dispersion relations of the polarization 
(\ref{omega}) and acoustic phonon (\ref{omegaac}) modes respectively, 
we can rewrite $\Sigma_\eta$ in the form~\cite{Khmel'nitskii71}
\begin{equation}
\label{sigmaac2}
\Sigma_\eta (r,T) = - 4\eta^2 \int \frac{d^3q}{(2\pi)^3} \, q^2
\Big \{ 
\frac{[n_B (\omega_p (q)) + \frac{1}{2}]}{\omega_p[\omega_a^2 - \omega^2_p]}
 + \frac{[n_B (\omega_a (q)) + \frac{1}{2}]}{\omega_a[\omega^2_p - \omega_a^2]}
\Big \}.
\end{equation}
At the quantum critical point, where $r=r_c$ and $T=0$,
the dispersion $\omega_p (q) = q$ and $n_B(\omega_p) = n_B(\omega_a) = 0$ so that
\begin{equation}
\label{sigmaac}
\Sigma_\eta (r_c,0) = - 4\eta^2 \int \frac{d^3q}{(2\pi)^3} \, 
\frac{1}{2\tilde{c}(\tilde{c}+1)q}.
\end{equation}
Using (\ref{tricky}), we write the gap function (as in IV A) as 
\begin{eqnarray}
\label{gapac}
\Delta^2 &=& \Omega_0^2 + \Delta_{\gamma_c}^2 + \Delta_\eta^2, \cr
\Delta_{\gamma_c}^{2} &=& 3 \gamma_c
\int\frac{d^{3}q}{(2\pi)^{3}}
\left(\frac{\big[ n_B(\omega_p (q)) + {\textstyle\frac{1}{2}
}\big]}{\omega_p} -\frac{1}{2q} \right), \cr
\Delta_\eta^2 &=&
-4\eta^2 \int
\frac{d^3q}{(2\pi)^3} \, q^2
\left( 
\frac{\big[ n_B(\omega_p (q)) + {\textstyle\frac{1}{2}} \big]}{\omega_p[\omega_a^2 - \omega^2_p]}
 + \frac{\big[ n_B(\omega_a (q)) + {\textstyle\frac{1}{2}} \big]}{\omega_a[\omega^2_p - \omega_a^2]}
- \frac{1}{2\tilde{c}[\tilde{c}+1]q^3}
\right), 
\end{eqnarray}
where $\Delta_{\gamma_c}^2$ has been already defined in (\ref{gapfunc}).

We emphasize that the $\gamma_c$ and $\eta$ terms in (\ref{gapac})
have opposite signs in their contribution  to the spectral gap $\Delta$. 
The negative coefficient of $\eta^2$ reflects the fact that it emerges from 
second-order perturbation theory; physically it is due to thermally enhanced virtual 
excitations caused by coupling between polarization TO and LA phonon modes.

\subsection{Deep in the Quantum Paraelectric Phase}

\figwidth=5in
\fg{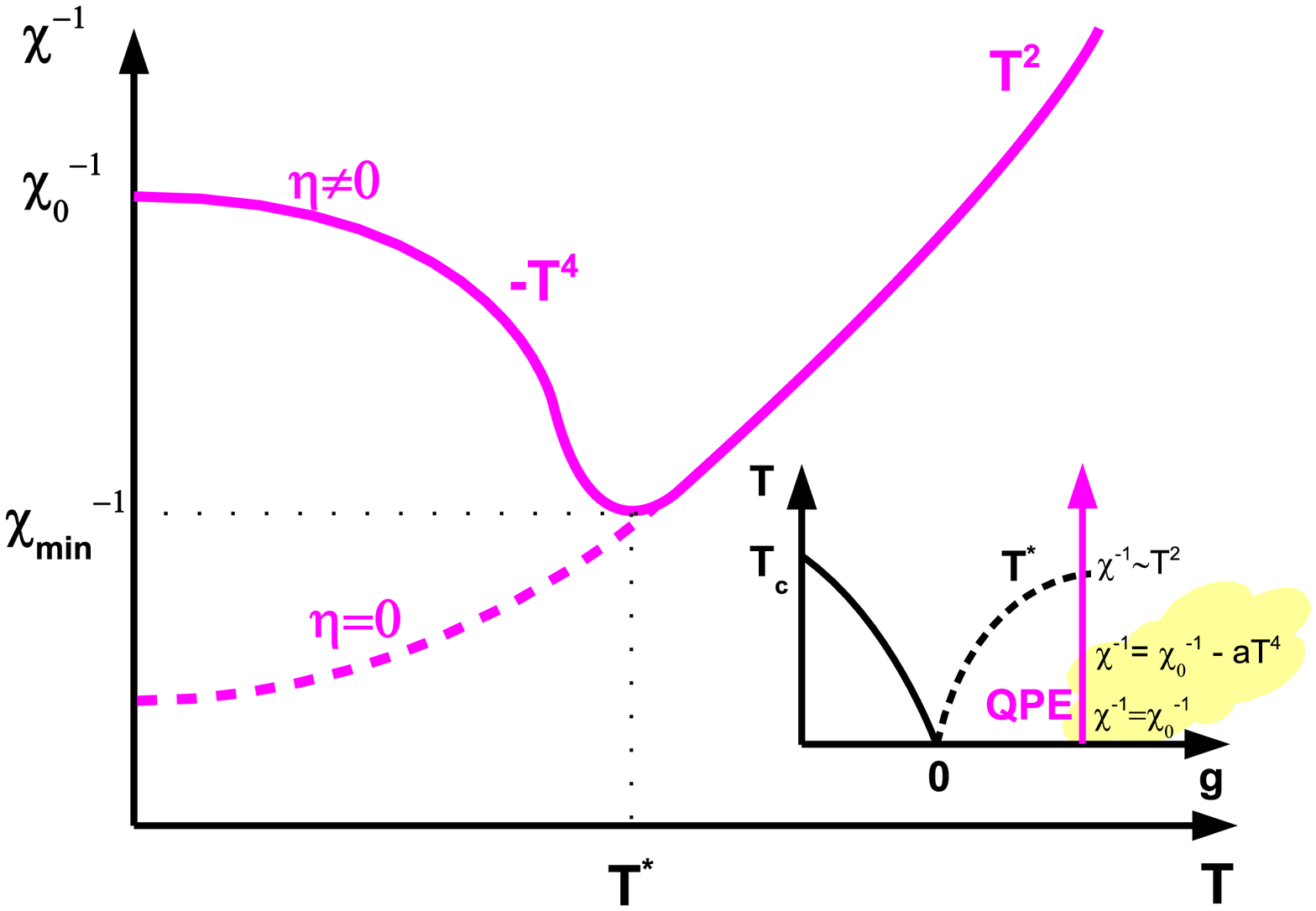}
{Schematic temperature-dependence of the static dielectric susceptibility where 
coupling to a long-wavelength acoustic phonon is included in
the calculation; inset indicates phase trajectory and region of 
corrections due to acoustic coupling deep in the QPE phase (yellow).}
{fig13}

Let us first explore the effect of the acoustic coupling $\eta$
deep in the QPE region of the phase diagram (see inset of Figure 9). 
Here $g>>0$ and $\Delta >> T \approx 0$.
In this regime, 
we write 
\begin{equation}
\label{chiacQPED>T}
\chi^{-1} = \Delta^2 = \Omega_0^2 + D(\Delta) - A(\eta) \, \frac{T^4}{\Delta^2}, 
\end{equation}
with  
\begin{equation}
A(\eta) = \frac{4\eta^2}{\tilde{c}} \int \frac{d^3 u}{(2\pi)^3} u \: n_B(\tilde{c}u),
\label{Aeta}
\end{equation}
where derivations of $A(\eta)$ and $D(\Delta)$ are presented in Appendix A;
for our purposes, the key point is to note that 
$\lim_{\Delta \rightarrow 0} D (\Delta) = 0$. 
Setting $A(\eta) \sim \eta^2 =0$, we recover a constant expression for $\chi$ 
as a function of temperature in the QPE phase, 
consistent with our previous derivations from Section IV. 
For $\eta \neq 0$,
the dielectric susceptibility acquires new low-temperature behavior.
The quartic temperature term in (\ref{chiacQPED>T}), $- A(\eta) \, \frac{T^4}{\Delta^2}$ , 
drives the inverse susceptibility at low temperatures;   
such a "bump" in the susceptibility (or "well" in the inverse susceptibility, see Fig. 9)
due to acoustic phonon coupling has been considered previously~\cite{Khmel'nitskii71}.
It is then natural to enquire whether a finite $\eta$ could eventually drive the inverse susceptibility 
to zero (or negative) values.   Here we show that this is not the case.
We start by looking for a solution of (\ref{chiacQPED>T}) with $\chi^{-1} = \Delta^2 = 0$, 
and show that such a solution does not exist.
Indeed at $\eta^2=0$, $\chi^{-1}$ in the QPE phase is nonzero as we saw in Section IV.
At $\eta^2\neq 0$, growth of last term in (\ref{chiacQPED>T}) exceeds all bounds and cannot
equate a constant term $\Omega_0^2$
(notice that $D(\Delta)|_{\Delta=0}=0$).
The inverse susceptibility therefore remains positive deep in the QPE phase with $\chi^{-1}_{min} \neq 0$.

We note that when the temperature increases so that
$\Delta \sim T$ and
we are no longer in the QPE phase (red in Fig. 7), 
we enter the "tornado" region of the QCP influence (blue in Fig. 7) 
where $\chi^{-1} \sim \Delta^2 \sim T^2$, as was shown in Section IV.
At this point the quadratic temperature-dependence dominates and coupling to the acoustic phonons 
becomes negligible; as a result
a turn-over in the inverse susceptibility from $-T^4$ to $+T^2$-dependence occurs (see Fig. 9).

\subsection{Quantum Critical Temperature-Dependent Dielectric Susceptibility}

We already know that there exists a classical phase transition  
at $T_c$ for $g<0$ and $\eta = 0$;
for $\eta \neq 0$ could this line of transitions enter the $g>0$ part of 
the phase diagram
and result in a reentrant quantum ferroelectric phase near the $g = 0$ QCP? 
In order to explore this possibility, we study the temperature-dependent susceptibility 
near the QCP (at $g=0$) and
find that unstable behavior is possible.
Next we follow the line of transitions, where 
$\chi^{-1} = \Delta^2 =0$
and show that its behavior is changed for $\eta > \eta_c$.

We begin with $\chi(T)$ in the vicinity in the quantum critical regime where $g=0$ (trajectory
2 in Figure 6);
here $\Omega_0^2 = g = 0$ and $q \sim \omega \sim T \gtrsim 0$
at low temperatures.
Taking $\eta = 0$, the spectral gap (\ref{gapac}) becomes
\begin{equation}
\label{gapgammaQC}
\Delta_{\gamma_c}^2 = \frac{3\gamma_c}{2\pi^2} \int dq \, q \, n_B( q/T)
\equiv \tilde{\alpha} \gamma_c T^2 = \frac{\gamma_c T^2}{4}
\end{equation}
and we recover the quadratic temperature dependence, 
$\chi^{-1}_{\gamma_c} = \Delta^2_{\gamma_c} \sim T^2$,
that was derived in Section IV B.

With $\eta \neq 0$, the $\eta$ contribution to the gap becomes
\begin{equation}
\label{gapetaQC}
\Delta_\eta^2 = - \frac{4\eta^2}{2\pi^2} 
\int \, dq \, \frac{q}{\tilde{c}^2 - 1} \left(n_B(q/T) -\frac{n_B(\tilde{c}q/T)}{\tilde{c}} 
\right) \equiv -\tilde{\beta}\eta^2 T^2.
\end{equation}
For both cases $\tilde{c} \lessgtr 1$, the expression under
the integral in (\ref{gapetaQC}) is positive (see Appendix B for specifics), 
which results in a negative coefficient for $\Delta_\eta^2$. 
Adding both $\gamma_c$ and $\eta$ terms in the gap equation (\ref{gapac}), we write
the expression for the dielectric susceptibility
\begin{equation}
\label{chiacQC}
\chi^{-1} = \Delta^2 = (\tilde{\alpha}\gamma_c - \tilde{\beta}\eta^2) T^2 = (\frac{\gamma_c}{4} - \tilde{\beta} \eta^2 ) T^2
\end{equation}
where $\tilde{\alpha}$ and $\tilde{\beta}$ are explicitly calculated in
Appendix C.
When the coefficient of $T^2$ is zero, namely when
\begin{equation}
\eta = \eta_c 
= \sqrt{\frac{\gamma_c}{4 \tilde{\beta}}}
=
\sqrt{\frac{3}{4}
\left(
\frac{\tilde{c}^{3}
(\tilde{c}^{2}-1)}{\tilde{c}^{3}-1} \right)
\gamma_{c}} 
\end{equation}
the phase boundary line 
($\chi^{-1} = 0$) becomes vertical in the approach to the QCP;
when $ \eta > \eta_c$, it ``meanders'' to the right leading to 
reentrant behavior. 

\figwidth=5in
\fg{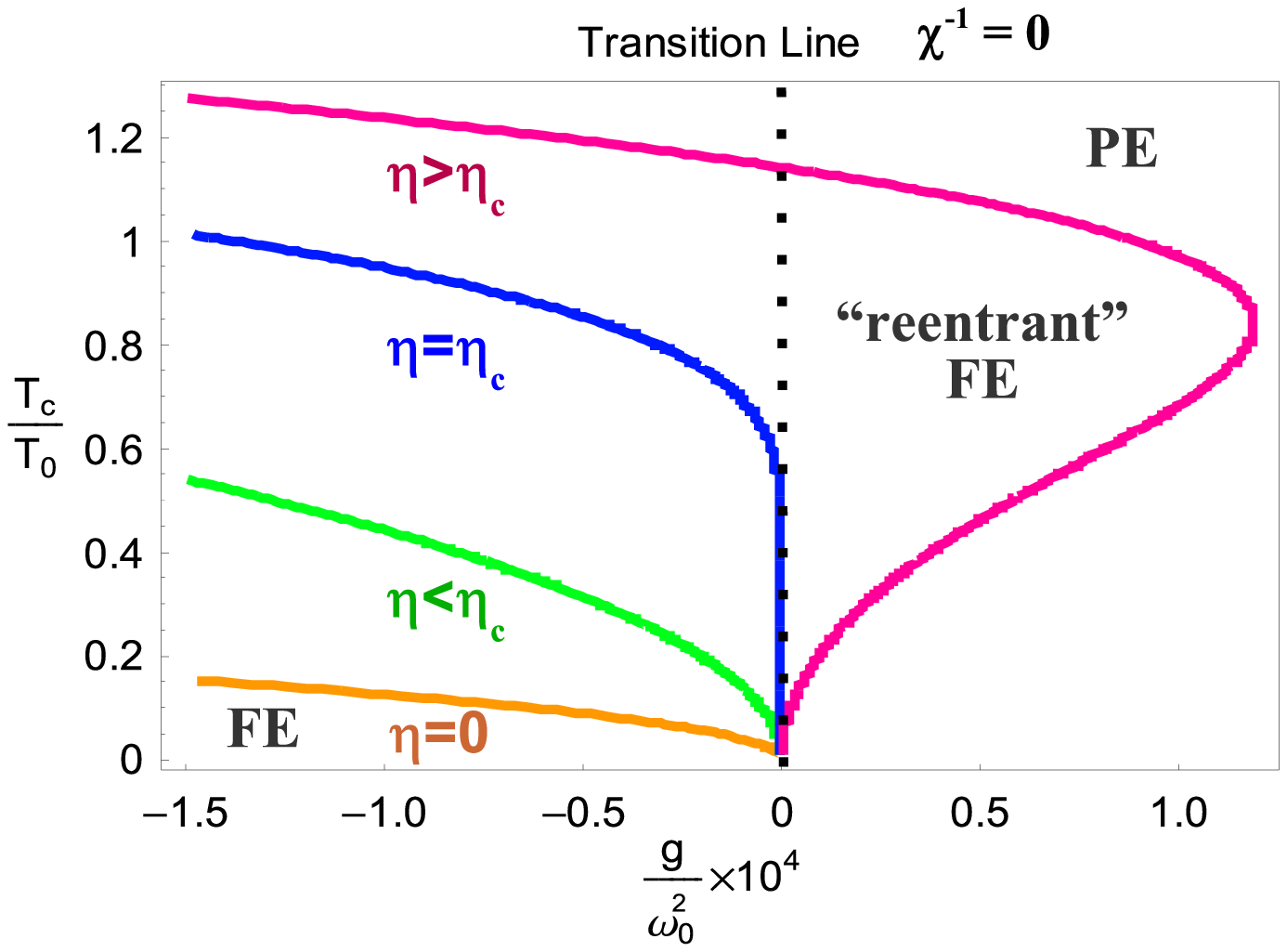}
{The transition line $T_c(g)$ for different values of $\eta$, the acoustic
coupling constant; for $\eta > \eta_c$ a reentrant quantum ferroelectric (FE)
phase emerges.
The phase boundaries result
from numerical solution of the gap equation ($\Delta (\eta \neq 0) = 0$)
discussed in the text; 
the parameters used here are:  $\gamma_c = 1$,$\tilde{c} = 0.9$, 
$\eta_c = 0.62$ and
$\{\eta > \eta_c, \eta < \eta_c \} = \{0.63,0.59\}$.}
{fig14}

\subsection{Details of the Phase Boundary ( $\chi^{-1} = 0$)}\label{details}

We now follow the phase transition line, defined by
$\chi^{-1}= 0 $ ($\Delta = 0$) out to finite temperatures.
From Section IV we know that there is a classical ferroelectric-paraelectric 
phase transtion at $g<0$ at Curie temperature $T_c (g)$; it
is depicted as a solid line in Fig. 6,
where
the dielectric susceptibility diverges, $\chi = \Delta^{-2} \to \infty$.
Our results in Section IV are for $\eta = 0$, and
we study the effect of $\eta > \eta_c > 0$ on this transition line.

To do this, we look for a solution to the gap equation (\ref{gapac}), when 
$\Delta (T_c,\eta) = 0$,
which yields the transition line $T_c (g,\eta)$. 
When the spectral gap closes, the dispersion relations of the 
TO soft polarization and the LA acoustic modes both become linear
($\omega_p(q)=q$ and $\omega_a(q)=\tilde{c} q$).
Inserting these values into (\ref{gapac}) and setting $\Delta = 0$, we obtain
\begin{eqnarray}
\label{gapline}
- 2 \pi^2 \Omega_0^2 \, &=& \, 3\gamma_c \int_0^{q_{max}} dq \, q \, n_B(q/T_c) 
  - \frac{4\eta^2}{(\tilde{c}^2- 1)} \int_0^{q_{max}} \, dq \, q 
  \Big\{n_B(q/T_c) -\frac{n_B(\tilde{c}q/T_c)}{\tilde{c}} \Big\} \cr
%
  &=& T_c^2 \left\{ 3\gamma_c \int_0^{u_{max}} du \, u \, n_B(u) 
  - \frac{4\eta^2}{(\tilde{c}^2-1)} \int_0^{u_{max}} du \, u 
  \Big\{n_B(u) -\frac{n_B(\tilde{c}u)}{\tilde{c}} \Big\} \right\}
\end{eqnarray}
for the equation determining $T_c(g)$.
At low temperatures, we note that we recover the scaling 
relation $\Omega_0^2 = g \sim T_c^2$ 
since both integrals become proportional to $T_c^2$
($u_{max} = \frac{q_{max}}{T_c}>>1$).
At high temperatures
$n_B(u)\approx \frac{1}{u}$, 
so the r.h.s. of (\ref{gapline}) becomes proportional to $T_c$,
and we recover the classical behavior $g \sim T_c$.

Figure 10 shows
the $T_c(g)$ transition line.
For $\eta > \eta_c \approx 0.6 $, the transition line ``wanders'' into the $g>0$ region, 
leading to a reentrant quantum ferroelectric phase.
Such reentrance suggests the possibility of nearby coexistence and
a line of first-order transitions ending in a tricritical point, but the
confirmation of this phase behavior requires a calculation beyond that presented
here and will be the topic of future work.

In order to make direct comparison with experiment, we must now restore dimensions to
our coupling constant and more generally to our Lagrangian (\ref{lagrangianac}).
We start by explicitly restoring all physical coefficients 
to the Lagrangian, as follows
\begin{eqnarray}\label{l}
\beta F& =&\int \frac{d^{3}\tilde{x}d\tau }{\hbar } L\cr
L& =&\frac{\alpha }{2}\left[(\partial_{\tau }\tilde{P})^{2} + c_{s}^{2} (\nabla
\tilde{P})^{2}\right]+
\frac{r_{D}}{2}\tilde{P}^{2}+\frac{\gamma_{D}}{4}\tilde{P}^{4}-\eta_{D}
(\nabla \tilde{{\phi }}) \tilde{P}^{2}+
\frac{\rho }{2}\left[(\partial_{\tau }{\tilde{\phi }} )^{2}+ c_{a}^{2} (\nabla \tilde{\phi})^{2}  \right]\nonumber
\\
\end{eqnarray}
where $c_s$ and $c_a$ are the soft optical and acoustic phonon
velocities respectively and where $\tilde{P}$ and $\tilde{\phi }$ are the
un-rescaled physical polarization and phonon displacement fields.
Then by writing
\begin{equation}
\frac{\tilde{x}}{c_{s}}= {x}, \qquad \frac{c_{s}^{3}\alpha }{\hbar }{\tilde{P}}^{2}=
{P^{2}}, \qquad \frac{c_{s}^{3}\rho }{\hbar }{\tilde{\phi}}^{2}={\phi }^{2}
\end{equation}
we obtain (\ref{lagrangianac}), the rescaled  Lagrangian,
\begin{eqnarray}\label{l}
\beta F &=& \int {d^{3}\tilde{x}d\tau }{} {\cal L}_{E} ({P},\phi )\cr
{\cal L}_{E} ( {P},\phi ) &= &\frac{1}{2}\left[(\partial_{\tau }{P})^{2} + (\nabla
{P})^{2}\right]+\frac{r}{2}{P}^{2} +\frac{\gamma}{4}{P}^{4}-\eta (\nabla {\phi}) {P}^{2}+
\frac{1 }{2}\left[(\partial_{\tau }{\phi} )^{2}+ \tilde{c}^{2} (\nabla {\phi})^{2}  \right]
\end{eqnarray}
where
\begin{equation}
\label{dimensionfulcs}
\frac{c_s^2}{\hbar} L = {\cal L}_{E} , \quad r = \frac{r_{D}}{\alpha }, \qquad \gamma =
 \frac{\hbar }{c_{s}^{3}\alpha^{2}}
\gamma_{D}, \qquad \eta = \frac{1}{\alpha c_{s}^{5/2}}\sqrt{\frac{\hbar}
{\rho }}\ \eta_{D}, \qquad \tilde{c}= \frac{c_{a}}{c_{s}}.
\end{equation}

In the dimensionless units used in this Section, we found that
\begin{equation}
\eta_{c}= \sqrt{\frac{\tilde{\alpha}}{\tilde{\beta }}\gamma_{c}}
=
\sqrt{\frac{3}{4}
\left(
\frac{\tilde{c}^{3}
(\tilde{c}+1)}{\tilde{c}^{2}+ \tilde{c}+1} \right)
\gamma_{c}}
\end{equation}
where $\tilde{\alpha }= \frac{1}{4}$ and  $\tilde{\beta }= \left(
\frac
{\tilde{c}^{2}+ \tilde{c}+1} {3\tilde{c}^{3}(\tilde{c}+1)}
\right)
$.
Using (\ref{dimensionfulcs}),
we can now rewrite this critical coupling in dimensionful terms as follows
\begin{eqnarray}\label{l}
\eta_{cD}&=& \sqrt{\frac{\rho }{\hbar }}{\alpha c_{s}^{5/2}}
\eta_{c}
\cr
&=& \sqrt{\frac{\rho }{\hbar }}{\alpha c_{s}^{5/2}} \sqrt{\frac{3}{4}\left(\frac{\tilde{c}^{3}
(\tilde{c}+1)}{\tilde{c}^{2}+ \tilde{c}+1} \right)}
\sqrt{ \frac{\hbar }{c_{s}^{3}\alpha^{2}}
\gamma_{cD}} 
\cr
&=& c_{s}\sqrt{{\rho }\gamma_{cD}}
\sqrt{\frac{3}{4}\left(\frac{\tilde{c}^{3}
(\tilde{c}+1)}{\tilde{c}^{2}+ \tilde{c}+1} \right)}.
\end{eqnarray}
For $SrTiO_3$, the acoustic~\cite{Schmising06,Bell63} and the soft-mode~\cite{Yamada69} 
velocities have been measured to be
$c_a \approx 8000 m/s$
and 
$c_s \approx 10000 m/s$ respectively so that 
$\tilde{c}=0.8$; 
the crystal mass density is $5.13 g/cm^3 = 5.13 \times 10^3 kg/m^3$. 
The value of $\gamma_c$ has been measured~\cite{Coleman06,Rowley07} 
using ferroelectric Arrott plots of $E/P$ vs $P^2$ to be 
$\epsilon_{0}\gamma_{cD}= 0.2 \quad  m^{4}/C^{2}$.
Inputting all these numbers and $\eta_c = 0.6$ into our dimensionful expression for $\eta_{cD}$,
we obtain
\begin{equation}
\label{etaDC}
\eta_{cD}^{STO}= 5.74 \times  10^{10 }Jm/C^{2}
\end{equation}
as the dimensionful critical coupling to be compared with experiment.

Next we estimate the experimental value of $\eta$ in $SrTiO_3$ as~\cite{Palova07}
$\eta^{STO} \approx \frac{Q}{s}$
where 
$Q$ and $s$ are the typical magnitudes of the electrostrictive constants and 
the elastic compliances~\cite{Pertsev98,Palova07} respectively; here we use
the values~\cite{Palova07} $Q = 0.05 \quad \frac{m^4}{C^2}$ and $s = 3 \times 10^{-12} \quad \frac{ms^2}{kg}$.
Thereofore we obtain
\begin{equation}
\eta_{STO} = 1.7 \times 10^{10} {Jm}/{C^2}
\end{equation}
so that
from our analysis we observe that $\eta_{STO}< \eta_{cD}^{STO}$
for the $SrTiO_3$ system.
However, there are two points of uncertainty here that we should emphasize:
(i) we use experimental values for  $SrTi^{16}O_3$
as they are not yet available for $SrTi^{18}O_3$; 
(ii) we use values of $Q$ and $s$ at room temperature, and
these quantities need to be determined at low temperatures. 
Despite the roughness of our estimate, 
it is reasonable to assume 
$\eta$ is not changed dramatically by the issues raised in (i) and (ii). 
We encourage further experimental investigations of $SrTi^{18}O_3$ at low temperatures
to clarify this situation.

\subsection{Translational-Invariance as Protection
against Damping Effects and Singular Interactions}\label{}

Our analysis of the effects of acoustic coupling has been limited to a
Hartree treatment of the leading self energy.  This approach neglects
two physical effects: \begin{itemize}
\item Damping, the process by which a soft mode phonon can
decay by the emission of an acoustic phonon 
\item The possibility of singular interactions induced by the exchange
of acoustic phonons 
\end{itemize}
Similar issues are of great importance in magnetic quantum phase transitions
in metals, where the coupling of the magnetization to the particle-hole
continuum of electrons introduces damping.\cite{Moriya73,Hertz76,Millis93}
For example, in the simplest
Hertz-Moriya treatment of a ferromagnetic quantum critical point,
damping by the electron gas gives rise to a quadratic Lagrangian of
the form
\begin{equation}\label{}
S_{M}= \sum_{q,\nu}
\biggl[
q^{2}+r + \frac{\vert \nu\vert}{q}
 \biggr]\vert M (q,\nu)\vert^{2}
\end{equation}
where the term linear in $|\nu|$ is a consequence of damping
by the particle-hole continuum.  This term plays a vital role in the quantum
critical behavior; by comparing the dimensions of the $q^{2}$ term
with the damping term, we see that $[\nu]\equiv q^{3}$, which means
that the temporal dimension scales as $z=3$ spatial dimensions under
the renormalization group. This has the effect of pushing the upper
critical dimension down from $4-1=3$ to $4-z=1$ dimensions. In
addition to this effect, the coupling to the electron-hole
continuum  also introduces non-local interactions
between the magnetization modes, casting doubt on the mapping
to a $\phi^{4}$ field theory.

Fortunately,  translational invariance protects the ferroelectric
against these difficulties. Translational invariance
guarantees that the soft mode can not couple directly to the
displacement of the lattice; instead it couples to the strain, the
gradient of the displacement, according to the interaction $H_{I}= -
\eta \nabla \phi  P^{2}$. When we integrate out the acoustic phonons,
the induced interaction between the soft-mode phonons takes the form
\begin{equation}\label{}
V (q,\nu) =  -4\eta^{2} \frac{q^2}{\nu^2+\tilde{c}^{2}q^{2}},
\end{equation}
where the numerator result from the coupling to the strain, rather
than the displacement.  The presence of the $q^{2}$ term in the numerator
removes the ``Coulomb-like'' $1/q^{2}$  divergence at small $q$,
protecting the soft mode interactions from the development of a
singular long range component.

A similar effect takes place with the damping.
To see this, we need to examine the imaginary part of 
the self-energy appearing in the Gaussian contribution to the 
action, (\ref{selfc}), 
\begin{equation}\label{gauss}
S _{G}= \sum_{q,\nu}\frac{1}{2}
\biggl[ \nu^2 + q^{2}+r + \Sigma_{\eta } (q)
\biggr ]\vert P (\vec{q},\nu)\vert^{2}.
\end{equation}
Damping results from the imaginary part
of self energy, $\Sigma''_{n} (q,\omega)$. To compute the damping, 
we generalize
$\Sigma_{\eta}$ given in (\ref{sigmaac2}) to finite frequency, obtaining 
\begin{equation}\label{}
\Sigma_{\eta } (q,z) = 
 2\eta^2 \int \frac{d^3k}{(2\pi)^3} \, k^2
\Big \{ 
\frac{[n_{p} + \frac{1}{2}]}{\omega_p[(z-\omega_{p})^{2}
- 
\omega_a^2 
]}
 + \frac{[n_{a}+ \frac{1}{2}]}{\omega_a[
 (z-\omega_a)^2-\omega^2_p 
]} + ( z\leftrightarrow -z)
\Big \},
\end{equation}
where we have used the short-hand $\omega_{a}\equiv \omega_{a} (k)$,
$\omega_{p} \equiv \omega_{p}(\vec{q}-\vec{k})$, $n_{a}= n_B (\omega_{a}
(k))$, $n_{p}= n_B (\omega_{p} (\vec{q}-\vec{k}))$. 
The imaginary part of
this expression at zero temperature, for positive $\nu$, is then given by
\begin{eqnarray}\label{l}
Im [\Sigma_{\eta } (q,\nu-i\delta )] &=&
 \pi\eta^2 \int \frac{d^3k}{(2\pi)^3} \,
\frac{k^{2}}{\omega_p \omega_{a}}
 \delta (\omega - \omega_{a}-\omega_{p}).
\end{eqnarray}
We can determine the small $q$, $\omega$ behavior of this damping rate
by simple dimensional analysis. The dimension of the right-hand side
is $[q^{5}]/[\omega^{2}]\sim q^{2}$, so the damping rate must have the form
\begin{equation}\label{l}
\Im \Sigma (q,\nu) \sim  \eta^2 \nu^2 F \left( \frac{q}{|\nu|},\frac{\Delta}{|\nu|} \right),
\end{equation}
where a more careful analysis of the integral reveals that 
$F \left( \frac{q}{|\nu|},\frac{\Delta}{|\nu|} \right)$ 
is not
singular at either small momentum or frequency. The most important
aspect of this result is that the scattering phase space grows
quadratically with frequency and momentum, so that it does not
dominate  over the other terms in the action (\ref{gauss}).
The scaling dimension of frequency remains the same as that of
momentum, and thus the upper-critical spatial dimension 
remains as $d=3$.

\section{Discussion}

\subsection{Logs, Dipolar Interactions and the Barrett Formula}

Before summarizing our results, let us briefly touch on a number
of topics closely related to our work which we have not yet discused;
more specifically they include logarithmic corrections in the upper critical dimension,
dipolar interactions and the use of the Barrett formula for quantum paraelectrics.
As we have already noted in Section VB, 
the polarization mode-mode interaction $\gamma_c$,
and coupling to the acoustic phonons $\eta$,
are both marginally relevant in the dimension of physical interest $d=3$.
Thus logarithmic corrections to the scaling relations (III)
have to be included; we have already seen their appearance in the 
expression for $\alpha_3$ in (\ref{alpha3}). 
The correction to scaling of the free energy  
near the classical ferro-paraelectric phase transition
in four dimensions is~\cite{Cardy96}
\begin{equation}
f_{cl}(t,\gamma_c) = f_0(t,\gamma_c) [1 + 9\gamma_c \, ln(t_0/t)]^{1/3},
\end{equation}
where $t= |\frac{T-T_c}{T_c}|$ is the reduced temperature,
$f_0(t,\gamma_c) = t^2 \Phi \left(\frac{E/E_0}{|t/t_0|^{3/2}} \right)$ 
is the scaling form of the free energy with a universal scaling function $\Phi$,
$t_0$ is the reduced Debye temperature for the soft mode (\ref{T0equation}) and
$\gamma_c$ is the polarization mode-mode coupling at QCP.
Since $\chi = \frac{\partial^2 f}{\partial E^2}|_{E=0}$,
we have
\begin{equation}
\chi = \chi_0 [1 + 9\gamma_c \, ln(t_0/t)]^{1/3},
\end{equation}
where $\chi_0 \sim t^{-1}$.
By applying the quantum-classical analogy (III), we write
at the upper critical dimension, $d_c^u=3$ ($d+z=4$; $z=1$),
\begin{equation}
f_{qm}(g,\gamma_c) = f_0(g,\gamma_c) [1 + 9\gamma_c \, ln(g_0/g)]^{1/3},
\end{equation}
where 
$g_0 \equiv \omega_0^2$ is the Debye frequency for the soft mode squared,
$f_0(g,\gamma_c)$ has the same form as before,
and
$g$ is the tuning parameter. 
By setting $\chi = \frac{\partial^2 f}{\partial E^2}|_{E=0}$,
the dielectric susceptibility becomes
\begin{equation}
\chi = \chi_0 [1 + 9\gamma_c \, ln(g_0/g)]^{1/3},
\label{glogscaling}
\end{equation}
where $\chi_0 \sim g^{-1} \sim T^{-2}$.
The temperature-dependence of $\chi$ with logarithmic corrections
is then found by making the subsitution $g \sim T^2$ in (\ref{glogscaling}),
and these results are identical to those found previously using
diagrammatic techniques\cite{Rechester71}.  An analogous
procedure can be used to find the logarithmic corrections
to other thermodynamic quantitites.  

We note that here we assume the upper critical (spatial) dimension
$d_c^u = 3$; however if we include uniaxial dipole-dipole interactions, 
we will have $d_c^u = 2$.  Basically this is because 
when all dipoles point in the (same) $z$-direction,
the TO polarization frequency (\ref{omega}) becomes~\cite{Larkin69}
\begin{equation}
\label{omegaqz}
\omega^2(q) = q^2 + \Delta^2 + \beta \frac{q_z^2}{q^2},
\end{equation}
where $\beta$ is a constant, and we derive (\ref{omegaqz}) in Appendix D.
We note that the last term of (\ref{omegaqz}) is specific 
to the uniaxial (e.g. tetragonal) case and is not present for
isotropic dipolar interactions.
Applying simple scaling, we obtain
\begin{equation}
\label{scaleqz}
\tilde{q}_{x(y)} = \frac{q_{x(y)}}{b}, \quad
\tilde{q_z} = \frac{q_z}{b^k},
\end{equation}
where the constants $b, \, b^k >1$ represent flow 
to the infrared (IR) limit of the QCP.
We show in Appendix D that in order for 
(\ref{omegaqz}) and (\ref{scaleqz}) to be satisfied
simultaneously, $k$ must equal $2$ so that
$q_z$ ``counts'' for effectively {\it two} dimensions
($d_{eff}^{space} = d + 1$),
so that for a quantum uniaxial ferroelectric
the total effective dimension is
$d_{eff} = d_{eff}^{space} + z = (d + 1) + z = d + 2$ with 
$d_c^u = 2$ since then we obtain $d_{eff} = 4$.

At this time, it is not known whether $SrTi^{18}O_3$ 
is cubic or tetragonal at low temperatures.  In any
case, we expect the samples under study to be structurally
multi-domain so that averaging over long length-scales
will make them effectively cubic; thus uniaxial
dipolar interactions do not need to be considered.
The observed $T^2$ behavior of $\chi$
in the vicinity of the QCP supports this contention
(i.e. $d_{eff}^{space} = 3$);
for $d_{eff}^{space} = 4$, a different $T$-dependence ($\chi^{-1} \sim T^3$)
is expected\cite{Schneider76} for a QPE so that a reexamination
of the underlying model would be necessary to match experiment. 
Until details of the samples are known, this situation
cannot be ascertained.  We note that such $T^2$ dependence of
the inverse susceptibility has also been observed\cite{Rytz80}
in mixed crystal ferroelectrics $KTa_{1-x}Nb_xO_3$
and $Ka_{1-y}Na_yTaO_3$ where uniaxial dipolar interactions
are not important, and we encourage further 
low-temperature studies of these systems.

A consistent discrepancy between 
the observed low-temperature dielectric susceptibility 
and the Barrett formula\cite{Barrett52} has been observed
in the quantum paraelectric phase.~\cite{Muller79,Rytz80}
Here we emphasize that the discrepancy occurs
when the system gets very close to the QCP;
thus it provides a measure of the tuning distance to the QCP.
Because the optical polarization mode softens as 
the system approaches the QCP, with
the gap vanishing completely here,
the momentum dependence in the dispersion relation (\ref{omega})
becomes important.
It is exactly for this reason that  the Barrett formula,
that assumes a constant dispersion relation, $\omega = \tilde{\omega}_0$,
breaks down close to the QCP.

The Barrett formula~\cite{Barrett52} works well deep in the QPE phase (V D),
where the gap is much bigger than temperature.
One such example is $KTaO_3$ (KTO),
which remains paraelectric down to zero temperature,
but in contrast to $SrTiO_3$ (STO) shows a much lower value of
the zero temperature dielectric susceptibility ($\chi_{KTO} \approx 4000$, 
$\chi_{STO} \approx 24000$)~\cite{Muller79,Akbarzadeh04}.
The closer the system is tuned to the QCP, the smaller is the
spectral gap and the bigger is the dielectric susceptibility.
Therefore, STO sits much closer to the QCP than KTO,
and indeed KTO shows a nice fit to the
Barrett formula~\cite{Akbarzadeh04}.
Notice that by plugging $\tilde{\omega}_0$ into (\ref{gapfunc}), 
we get the Barrett expression,
\begin{eqnarray}
\chi^{-1} = \Delta^2 &=& \Omega_0^2 + \frac{3\gamma_c}{4\pi^2} 
\left(
\frac{\coth(\tilde{\omega}_0/2T)}{\tilde{\omega}_0}\frac{q_{max}^3}{3} - \frac{q_{max}^2}{2}
\right) \cr
&=& \frac{1}{M} \left( \frac{T_1}{2} \coth(T_1/2T) - T_0 \right),
\end{eqnarray}
where $T_1 \equiv \tilde{\omega}_0$, and $M$ and $T_0$ are
fitting constants.

\subsection{Summary and Open Questions}

Let us now summarize the main results of the paper.  Here
our aim has been to characterize the finite-temperature
properties of a material close to its quantum ferroelectric
critical point; we have rederived and extended previous
theoretical results using scaling methods and self-consistent
Hartree theory. In the process we have made an analogy
between temperature as a boundary effect in time
and the Casimir effect, and have used this to shed light
on both problems.  Using simple finite-size scaling, we
have presented straightforward derivations of finite-temperature
observables for direct comparison with experiment, and
our approach has yielded a scaling form 
$\chi(\omega) = \frac{1}{\omega} F(\frac{\omega}{T})$
which serves as an additional probe of $T_0$, the soft-mode Debye
temperature-scale where we expect crossover between Curie ($T$)
and Quantum Critical ($T^2$) behavior in $\chi^{-1}$.
We emphasize that this scaling method is useful
in this system where $z$ is low ($z = 1$); otherwise
if $z$ is higher, the system is usually well above its
upper critical dimension where this approach is
inappropriate. Next we've used self-consistent Hartree methods to
determine the $T-g$ phase diagram
and the crossover between classical and quantum behavior.
In particular we see the influence of the quantum critical
point on the susceptibility at finite temperatures, and
we can put in materials parameters to determine the
size of its basin of attraction.  Finally we include
coupling to an acoustic phonon and find that it affects
the transition line; for such couplings greater than
a threshhold strength there is a reentrant quantum
ferroelectric phase.  

Naturally these results suggest a number of open questions
and here we list a few:

\begin{itemize}

\item The presence of a reentrant phase suggests the possibility
of nearby phase coexistence, a tricritical point and a line
of first order transitions.  This is a particularly appealing
scenario given that recent experiments\cite{Taniguchi07} suggest coexistence
of QPE and QFE in $SrTi^{18}O_3$ and is a topic we plan to pursue
shortly.

\item If indeed there is a tricritical point and a line of first-order
phase transitions, could there also be a metaelectric critical
point in the $g-E$ plane analogous to the metamagnetic situation\cite{Millis02,Gegenwart08}
in some metallic systems? There is indication that an analogous metaelectric
critical point occurs in a multiferroic system,\cite{Kim08} so this is a question
driven by recent experiment.

\item What happens when we add spins to a system near its
quantum ferroelectric critical point? Would the resulting
multiferroic have particularly distinctive properties?

\item Similarly what type of behavior do we expect if we dope
a quantum parelectric in the vicinity of a QCP?  There is by now
an extensive body evidence that 
electronically mediated superconductivity 
is driven by the vicinity to a magnetic quantum critical point, 
phenomenon of ``avoided
criticality", whereby superconductivity in the vicinity of a naked
magnetic quantum critical point\cite{mathur,avoided}.  In such systems,
the metallic transport properties develop strange metallic properties that
have been termed ``non-Fermi liquid behavior"\cite{piers,rosch}. This
raises the important question, as to what, if any, is the ferroelectric
counterpart to this behavior? In particular
- how does the presence of a soft mode affect the semi-metallic properties
of a doped quantum critical ferro-electric, and does a doped
ferroelectric quantum critical point also develop superconductivity via
the mechanism of avoided criticality?

\end{itemize}

We believe that we have only begun to explore the rich physics associated with
the quantum ferroelectric critical point, a simple setting for studying
many issues associated with quantum criticality that emerge in much more
complex materials.  Furthermore the possibility of detailed interplay between
theory and experiment is very encouraging.

\section{Acknowledgments}
We thank D. Khmelnitskii, G.G. Lonzarich, S.E. Rowley, S.S. Saxena and J.F. Scott
for detailed discussions. We are particularly grateful to H. Chamati and to N.S. Tonchev for pointing us to an error in the numerical analysis of equation (41) in an earlier
 version of this paper. We also acknowledge financial support from
the National Science Foundation NSF-DMR 0645461 (L.Palova),
NSF-NIRT-ECS-0608842 (P. Chandra) and the Department of Energy, 
grant DE-FE02-00ER45790 (P. Coleman). 

\section{Appendix A:  $D(\Delta)$ and $A(\eta)$}

We derive expressions for $D(\Delta)$ and $A(\eta)$ (\ref{Aeta})
using the gap equation (\ref{gapac}) deep in the QPE region (D),
where $g>>0$ and $\Delta>>T\approx 0$.
Collecting all ``$\frac{1}{2}$''-terms under integrals of
$\Delta_{\gamma_c}^2$ and $\Delta_\eta^2$ in (\ref{gapac}), we obtain the
expression for $D(\Delta)$,
\begin{eqnarray}
D(\Delta) &\equiv& \frac{3}{2}\gamma_c \int \frac{d^3q}{(2\pi)^3} \left(\frac{1}{\omega_p} - \frac{1}{q}\right)
-2\eta^2 \int \frac{d^3q}{(2\pi)^3} q^2 \left( \frac{1}{\omega_a [\omega_p^2 - \omega_a^2]} - 
\frac{1}{q^3 \tilde{c} [1 - \tilde{c}^2] } \right) \cr
&-& 2\eta^2 \int \frac{d^3q}{(2\pi)^3} q^2 \left( \frac{1}{\omega_p [\omega_a^2 - \omega_p^2]} - 
\frac{1}{q^3 [\tilde{c}^2 - 1]}\right) 
\equiv \frac{3\gamma_c}{4\pi^2} I_1 - \frac{\eta^2}{\tilde{c}\pi^2} I_2 - \frac{\eta^2}{\pi^2} I_3, \cr
I_1 &=& \int_0^{q_{max}} dq \, q^2 \left( \frac{1}{\sqrt{\Delta^2+ q^2}} - \frac{1}{ q} \right), \cr
I_2 &=& \int_0^{q_{max}} dq \, q^3 \left( \frac{1}{\Delta^2 + q^2[1 -\tilde{c}^2]} - \frac{1}{q^2[1 -\tilde{c}^2]}\right)
, \cr
I_3 &=& \int_0^{q_{max}} dq \, q^4 \left( \frac{1}{\sqrt{\Delta^2 + q^2}[-\Delta^2 + q^2[\tilde{c}^2-1]} 
- \frac{1}{q^3[\tilde{c}^2-1]}\right).
\end{eqnarray}
Notice that lim$_{\Delta \to 0} D(\Delta) = 0$, since all three integrals $I_1$, $I_2$ and $I_3$ become zero
at zero gap.
We split the integrals $I_i$ ($i=1$,$2$,$3$) into two parts, 
$I_i = \int_0^{n\Delta} + \int_{n\Delta}^{q_{max}}$,
where $n\Delta>>\Delta$.
Since $q>>\Delta$ in the second integral part, we neglect its $\Delta$ dependence 
and get a zero contribution.
Thus, only the first integral part contributes, and $D(\Delta)$
becomes a function of $\Delta$ only, with no temperature dependence.

Next we show that the second Bose-Einstein $n_B(\omega_a(q))$ term under the 
integral of $\Delta_\eta^2$ in (\ref{gapac}) results in the 
form $A(\eta)$ in equation (\ref{Aeta}), 
\begin{eqnarray}
\label{Aetaderiv}
-4\eta^2 \int \frac{d^3q}{(2\pi)^3} \, q^2 
\frac{n_B(\omega_a (q))}{\omega_a [\omega_p^2 - \omega_a^2]} 
&=& -4\eta^2 \int \frac{d^3q}{(2\pi)^3} \, q^2 
\frac{n_B(\omega_a (q))}{\tilde{c} q [\Delta^2 + q^2 (1 - \tilde{c}^2)]} \cr
&\approx& -\frac{4\eta^2}{\tilde{c}}\frac{T^4}{\Delta^2} \int \frac{d^3 u}{(2\pi)^3} u \: n_B(\tilde{c}u)
\equiv - A(\eta) \frac{T^4}{\Delta^2},
\end{eqnarray}
where $u=q/T$.
Notice that we approximate $\Delta^2 >> q^2(1 -\tilde{c}^2)$ in the second line of (\ref{Aetaderiv}).
For low momenta, this is indeed the case.
For large momenta, $q>> \Delta >> T \approx 0$, we neglect $\Delta$  
in (\ref{Aetaderiv}) and the integral becomes
\begin{equation}
\label{Aetaqlarge}
-\frac{4\eta^2}{2\pi^2\tilde{c}(1 -\tilde{c}^2)} \int dq \, q n_B(\tilde{c}q).
\end{equation}
In the limit $q>>T$, $n_B(\tilde{c}q)\approx e^{-\tilde{c}q/T}$ and (\ref{Aetaqlarge}) becomes 
exponentially small ($\sim T^2 e^{-\tilde{c}q/T}$) and can be neglected.
Similarly, we neglect the rest of the terms in the gap function (\ref{gapac})
with Bose-Einstein thermal distribution $n_B(\omega_p(q))$.
Deep in the QPE phase $\Delta >> T$, so that $n_B(\omega_p(q))\approx e^{-\Delta/T}$
at low momenta, or $n_B(\omega_p(q))\approx e^{-q_{large}/T}$ at large momenta.
In both cases $\Delta, q_{large} >> T$, the integrals containing $n_B(\omega_p(q))$
become exponentially small and so are negligible.

%
%

\section{Appendix B: Integral (\ref{gapetaQC}) is positive  for $\tilde{c} \lessgtr 1$}

We also show that the expression under the integral in (\ref{gapetaQC})
is positive for the two cases, $\tilde{c} \lessgtr 1$. First, assuming
that $\tilde{c} < 1 $, $\tilde{c} q <  q$ (positive $q$'s) and 
$n_B(\tilde{c} q/T) > n_B( q/T)$ we write
\begin{eqnarray}
\Big\{n_B(q/T) -\frac{n_B(\tilde{c}q/T)}{\tilde{c}} \Big\} 
\frac{1}{\tilde{c}^2-1} &>& (1 -\frac{1}{\tilde{c}}) \, n_B(\tilde{c}q/T)
\frac{1}{\tilde{c}^2-1} \cr
 &=& \frac{1}{\tilde{c} (\tilde{c} +1)} n_B(\tilde{c}q/T) \geq 0,
\end{eqnarray}
which we note is positive.
Similarly, for $\tilde{c}> 1$, $\tilde{c} q >  q$ and $n_B(\tilde{c} q/T) < n_B (q/T)$,
we write
\begin{eqnarray}
\Big\{n_B(q/T) - \frac{n_B(\tilde{c}q/T)}{\tilde{c}} \Big\} 
\frac{1}{\tilde{c}^2-1} &>& (1 - \frac{1}{\tilde{c}}) \, n_B(q/T)
\frac{1}{\tilde{c}^2-1} \cr
 &=& \frac{1}{\tilde{c} (\tilde{c} +1)} n_B( q/T) \geq 0.
\end{eqnarray}
which is also positive.  Therefore the integral in (\ref{gapetaQC})
is positive in both cases.

\section{Appendix C: $\tilde{\alpha}$ and $\tilde{\beta}$ are constants}

To evaluate the quantities  $\tilde{\alpha}$ and $\tilde{\beta}$ in
(\ref{gapgammaQC}) and (\ref{gapetaQC}),  we make a change of variables
to $u= q/T$,  and $u= \tilde{c} q/T$ respectively. The expressions for these two constants
then become 
\begin{eqnarray}
\label{alphabetaderiv}
\tilde{\alpha} &=& \frac{3}{2\pi^2} \int_0^{q_{max}/T} du \, u
\, n_B(u) = \frac{1}{4},
\cr
\tilde{\beta} &=& 
\frac{2}{\pi^2 (\tilde{c}^2 -1)} \left[
\int_0^{q_{max}/T} 
- 
\frac{1}{\tilde{c}^3}\int_0^{\tilde{c} q_{max}/T} 
\right]  du u n_B(u)
=
\frac{1}{3 (\tilde{c}^2 -1)} \left(1-\frac{1}{\tilde{c}^3}
\right),
\end{eqnarray}
where we have taken the limits of integration to infinity and used the
result \mbox{$\int_{0}^{\infty }du\ u n_B (u)=\frac{\pi^{2}}{6}$}.

\section{Appendix D: Dipole-dipole interactions in uniaxial ferroelectrics}

The interaction energy between two dipoles $\vec p_i$ and $\vec p_j$ siting
on two sites $\vec r_i$ and $\vec r_j$ respectively is
\begin{equation}
\label{dipoleWij}
W_{ij} (\vec r) = \frac{\vec p_i \cdot \vec p_j - 3(\vec n \cdot \vec p_i)
(\vec n \cdot \vec p_j)}{4\pi\epsilon_0 |\vec r|^3},
\end{equation}
where $\vec n$ is a unit vector in the direction of the vector 
$\vec r \equiv \vec r_j - \vec r_i$.
From (\ref{dipoleWij}), we find the total dipole-dipole interaction
potential to be
\begin{equation}
\label{dipoleW}
W (\vec r) = \frac{1}{4\pi \epsilon_0} \sum_{i,j,a,b} p_i^a p_j^b \left( \frac{\delta_{ab}}{r^3}
- \frac{3 r^a r^b}{r^5} \right),
\end{equation}
where $r \equiv |\vec r|$, and $a,b$ label vector coordinates.
After we perform a Fourier transform, the interaction potential becomes
\begin{equation}
\label{FourierW}
W (\vec q) = \frac{1}{\epsilon_0} \sum_{a,b} p^a_{\vec q} \, p^b_{-\vec q} \, \frac{q_a q_b}{q^2},
\end{equation}
where $q \equiv |\vec q|$ refers to the momentum-dependence of $W(\vec q)$.
Assuming that all dipoles point in the same($z$)-direction in the 
uniaxial case, we find that the dipole potential 
\begin{equation}
W(\vec q) \sim \frac{q_z^2}{q^2}.
\end{equation}
$W(\vec q)$ contributes to Lagrangian (\ref{lagrangianac}),
$L_E [P,\Phi] \to L_E [P,\Phi] + W$,
so that the TO polarization frequency (\ref{omega}) then reads~\cite{Larkin69}
\begin{equation}
\omega^2(q) = c_s^2 q^2 + \Delta^2 + \beta \frac{q_z^2}{q^2},
\end{equation}
where we introduce constant of proportionality $\beta$.

We show that (\ref{omegaqz}) and (\ref{scaleqz}) conditon $k=2$.
Let us assume that $k>1$. Then
\begin{eqnarray}
\tilde{q}^2 &=& \tilde{q_x}^2 + \tilde{q_y}^2 + \tilde{q_z}^2 =
\frac{q_x^2 + q_y^2}{b^2} + \frac{q_z^2}{({b^2})^k} \approx \frac{q^2}{b^2}, \cr
\frac{\tilde{q}_z^2}{\tilde{q}^2} &\approx& b^{2-2k} \, \frac{q_z^2}{q^2}.
\end{eqnarray}
Since we also rescale frequency $\omega (q)$ (\ref{omegaqz}) 
by a constant,
expressions,
$\tilde{q}^2$ and $\frac{\tilde{q_z}^2}{\tilde{q}^2}$,
are to be proportional. This leads then to the condition
\begin{equation}
k=2.
\end{equation}

%
\def\refname{Bibliography}

\end{document}